\documentclass[preprint2]{aastex}

\def\a{\alpha}
\def\b{\beta}
\def\d{\ell}
\def\e{\epsilon}

\def\L{\Lambda}
\def\r{\rho}
\def\g{\gamma}

\def\s{\sigma}
\def\ra{\rightarrow}
\def\Ra{\Rightarrow}

\def\bm#1{\mbox{\boldmath{$#1$}}}

\begin{document}

\title{Halos and voids in a multifractal model of cosmic structure}

\author{Jos\'e Gaite}
\affil{Instituto de Matem{\'a}ticas y F{\'\i}sica Fundamental,
CSIC, Serrano 113bis, 28006 Madrid, Spain}

\date{29 November 2006}


\begin{abstract}
On the one hand, the large scale structure of matter is arguably scale
invariant, and, on the other hand, halos and voids are recognized as
prominent features of that structure.  To unify both approaches, we
propose to model the dark matter distribution as a set of fractal
distributions of halos of different kinds.  This model relies on the
concept of multifractal as the most general scaling distribution and
on a plausible notion of halo as a {\em singular} mass concentration
in a multifractal.  Voids arise as complementary to halos, namely, as
formed by {\em regular} mass depletions.  To provide halos with
definite size and masses, we coarse-grain the dark matter distribution
using a natural length derived from the lower scaling limit.  This
allows us to relate the {\em halo mass function} to the {\em
multifractal spectrum}.  Hence, we find that a log-normal model of the
mass distribution nicely fits in this picture and, moreover, the
Press-Schechter mass function can be recovered as a {\em bifractal}
limit.

To support our model of fractal distributions of halos, we perform a
numerical study of the distribution produced in cosmological $N$-body
simulations.  In the Virgo $\L$-CDM GIF2 simulation, we indeed find
fractal distributions of halos with various dimensions and a halo mass
function of bifractal type. However, this mass function is just beyond
the Press-Schechter's range, and we interpret it instead as caused by
the {\em undersampling} of the distribution at the scale of halos, due
to discretization.
\end{abstract}


\section{Introduction}

Current research on the large scale structure of matter is concerned
with two fundamental concepts: halo and void.  The concept of halo
applies to the dark matter distribution and is meant to define the
basic unit of structure.  Therefore, dark matter halos have
theoretical rather than observational basis. In particular, the
concept of halo is useful in the analysis of $N$-body cosmological
simulations. In contrast, the concept of void arose in observations of
the galaxy distribution, although, recently, it has also been applied
to the dark matter distribution and gained some theoretical basis.

On a more fundamental level, the principle of scale invariance has
been often applied to the study of the large scale structure of
matter.  Scale invariance is a general symmetry, with applications in
various branches of physics and other sciences. Its application in
cosmology is old, dating back to hierarchical models of the Universe
devised before the discovery of General Relativity (Mandelbrot,
1977). It became part of standard cosmology when the analysis of
galaxy catalogues proved that the two-point correlation function is a
power law on scales of several Mpc. A recent review of various ideas
and models in cosmology based on scaling laws is given by Jones et al
(2004). Of particular importance are fractal models (Sylos Labini,
Montuori \& Pietronero, 1998). Multifractal models (Parisi \& Frisch,
1985; Halsey et al, 1986) are an improvement of fractal models and
were introduced in cosmology by Pietronero (1987), Jones et al (1988)
and Balian \& Schaeffer (1988).  Multifractal analyses of large-scale
structure were initially applied to the distribution of galaxies
(Dom{\'\i}nguez-Tenreiro \& Mart{\'\i}nez, 1989; Mart{\'\i}nez, Jones,
Dom{\'\i}nguez-Tenreiro \& van de Weygaert, 1990) and later applied to
$N$-body simulations (Colombi, Bouchet \& Schaeffer, 1992; Valdarnini,
Borgani \& Provenzale, 1992; Yepes, Dom{\'\i}nguez-Tenreiro \&
Couchman, 1992).

We shall take scale invariance of the dark matter distribution in the
{\em nonlinear} regime as our working hypothesis. This hypothesis
leads essentially to a multifractal model (with transition to
homogeneity). Since the linear regime departs from the initial scale
invariance (Harrison-Zeldovich spectrum), the initial conditions for
nonlinear evolution are not scale invariant. However, the dynamical
equations for cold dark matter are scale invariant. Therefore, it is
reasonable to assume that the dynamics is driven to a scale-invariant
attractor that is independent of the initial conditions, as is normal
in nonlinear systems. 

The power-law form of the galaxy-galaxy correlations can be taken as
evidence of scale invariance, although there may be deviations from
pure power laws.  In particular, Zehavi et al (2004) find small but
systematic deviations in the projected correlation function of a
sample from the Sloan Digital Sky Survey. They relate those deviations
to the deviations from a power-law correlation function found by
Jenkins et al (1998) in $\L$-CDM simulations. Hence, they interpret
them as caused by the transition from a two-halo regime on large
scales to a one-halo regime on small scales.  We shall discuss an
alternative interpretation at the end of this paper, after we have
introduced and explained the multifractal model of halos. Our
interpretation is fully consistent with scale invariance.

Halo models of large-scale structure are now well developed (Cooray \&
Sheth, 2002).  They assume that the full dark matter distribution is
composed of two parts: the dark matter distribution within halos and
the distribution of halos themselves.  The matter distribution within
halos is considered to be in virial equilibrium.  This inner
distribution has received much attention.  Usually, it is assumed to
be relatively smooth, and it is useful to define the average
``profile'' of a single halo.  The first approximation is a power-law
profile (Peebles, 1974; Sheth \& Jain, 1997; Murante et al, 1997).
However, more refined profile models involve a crossover between two
power laws (Cooray \& Sheth, 2002).

Regarding the distribution of halos, halo centers have been
distributed randomly in some models (Peebles, 1974; McClelland \&
Silk, 1977; Sheth \& Jain, 1997; Murante et al, 1997) but clustering
in the distribution of halos has also been considered (Mo \& White,
1996; Sheth \& Tormen, 1999).  Clustering of halos is to be expected
and, in particular, the clustering of halos can be of fractal type.
In fact, a power-law halo profile and a fractal distribution of halos
are just two aspects of a new formulation of the multifractal model
(Gaite, 2005-A).  In this formulation, a general multifractal is
considered as a set of fractal distributions of halos, such that
similar halos have a fractal distribution with a given dimension.

However, the separation of the full matter distribution between a
distribution within halos and a distribution of halos is not
meaningful if the full distribution is scale invariant.  A power-law
profile actually prevents one from assigning definite sizes or masses
to halos, which must be associated, in principle, with {\em
point-like} mass singularities (Gaite, 2005-A).  But the dark matter
distribution is not scale invariant on ever decreasing scales.  Thus,
the breaking of scale invariance induces a change in the nature of the
matter distribution that allows us to assign halos a definite size.

More in general, we will see that the fine structure intrinsic to
mathematical multifractals on ever decreasing scales gives rise to
some counter-intuitive and even paradoxical geometrical notions. Some
of these notions are relevant to describe the dark matter
distribution, in spite of its not being scale invariant on ever
decreasing scales.

Thus, our definition of halo differs from the usual definition: we
understand that a natural definition of halo in a multifractal model
with a natural coarse-graining scale assigns halos equal size, namely,
the size of that scale.  A similar definition, namely, based on a
uniform size, was proposed by Vergassola et al (1994) to define a mass
function for the structure produced in the adhesion model.  In
contrast, the most popular definition of halo is based on the {\em
spherical collapse} formalism (Mo \& White, 1996; Sheth \& Jain, 1997,
Cooray \& Sheth, 2002), and it assumes that halos are given by a
density-contrast threshold ($\delta \r/\r = 178$). This definition of
halo constitutes the basis for the halo finders that use the
``friends-of-friends'' (FoF) technique (Davis et al,
1985). Nevertheless, Valageas, Lacey \& Schaeffer (2000) have shown
that both choices, namely, halos given by a density condition or
equal-size halos, can be considered adequate from a theoretical point
of view.  Our multifractal model of halos agrees with their
conclusion, for its crucial feature is the pattern in which the mass
concentrates rather than the division of mass among individual halos,
which can be made in various manners. Of course, the simplest manner
is to take equal-size halos.

The notion of void as basic element of the large scale structure is
now well established, but its definition is more imprecise than the
definition of halo. The original definition of void as a totally empty
region, suitable for the galaxy distribution, is now being replaced
with definitions that allow for partial emptiness, after having been
admitted that dark matter and even low luminosity galaxies are present
in the initial voids.  In fact, recently, Shandarin, Sheth \& Sahni
(2004) have defined void as an {\em underdense connected} region,
such that voids are complementary to superclusters. We propose here a
definition that is related with the definition of Shandarin, Sheth \&
Sahni (2004), but which hinges on the nonlinear multifractal nature of
the distribution, unlike theirs.

We begin with a brief review of of multifractals, focusing on the
concept of mass concentrations (singularities) and the realization of
scale invariance. We introduce the key concept of multifractal
spectrum and, in particular, the bifractal spectrum, as well as the
method of calculation of the spectrum from correlation moments.  Halos
can be identified with singular mass concentrations, with power-law
profile, whereas voids can be identified with regular mass depletions.
Useful computational notions of halo and void result when we consider
a (natural) coarse-graining scale.  This leads us to a general study
of the coarse-grained probability distribution function and its
connection with the lognormal model.  Further connection with the
Press-Schechter mass function is also possible.  Then we study what
implications a multifractal model has for the distribution of halos
(as we have defined them).  Finally, we test all these notions against
cosmological $N$-body simulations.

\section{Multifractals and mass concentrations}

Multifractal measures appear frequently in physics.  In particular,
they appear frequently as attractors of dynamical systems (Halsey et
al, 1986; Falconer, 2003). In dynamical systems, the natural measure
of some region represents the portion of time that the system spends
in it. For many systems, one indeed finds that, at long times, the
trajectory is confined to a small region of the available space (an
attractor), but the time that it spends in any subregion of it is very
variable. In other words, the measure is very irregular and this
non-uniformity encodes further structure that defines the attractor.

The formation of large scale structure is given by the cosmological
Newton-Poisson equations (Peebles, 1980), which constitute a dynamical
system. Therefore, to justify the relevance of multifractal analysis
in the description of large scale structure, we may adopt the natural
view that this structure is just an attractor of that many-dimensional
dynamical system. Or we may argue that we expect a scale-invariant
structure, but in which the realization of scale invariance is
non-trivial, as in general multifractals. At any rate, the final
justification must rest on a successful comparison with observational
data or results of cosmological simulations.

As said above, multifractal measures represent mass distributions
spread according to highly irregular patterns, that is, with mass
concentrations of very different magnitude (Falconer, 2003). This
magnitude is referred to as ``strength'' and defined by the local
dimension $\a$ (also called Lipschitz-H\"older exponent in the
mathematical literature); namely, 
\begin{equation}
\a({\bm x}) = \lim_{r\ra 0}\frac{\log m[B({\bm x},r)]}{\log r},
\label{a}
\end{equation}
where $m[B({\bm x},r)]$ is the mass in the ball of radius $r$ centered
on the point ${\bm x}$.  We also write, informally, $m[B({\bm x},r)]
\sim r^{\a(}\mbox{\boldmath{$^x$}}^{)}$.  We only consider points
${\bm x}$ such that $m[B({\bm x},r)]$ does not vanish for any $r > 0$,
to prevent the divergence of $\a({\bm x})$. These points form the {\em
support} of the measure.

In a regular mass distribution, $\a = 3$ (constant). So a multifractal
is a singular mass distribution and its mass concentrations are also
called singularities (of various strengths).  On the other hand, an
ordinary self-similar fractal can be considered endowed with a uniform
mass distribution over it, such that $\a < 3$ is the {\em constant}
fractal (Hausdorff-Besicovitch) dimension. Thus, in the context of
multifractals, ordinary self-similar fractals are called {\em
monofractals} (or {\em unifractals}). Full-fledged multifractals
possess a range of exponents $\a$ (or strenghts), namely, $0 \leq
\a_{\rm min} \leq \a \leq \a_{\rm max}$. Every set of points in which
$\a$ takes a definite value is a fractal set (in the sense of having
non-trivial fractal dimension). Therefore, a multifractal can be
considered as a set of fractals containing the various mass
concentrations.  The {\em multifractal spectrum} $f(\a)$ is the
function that gives the fractal dimension of the set of points with
exponent $\a$.

Regarding multifractal spectra, we should mention that there are cases
in which $\a$ is not strictly constant but which cannot be considered
full-fledged multifractals. An example is a finite number of
(isolated) concentrations of different strength. This is a regular
distribution except for the finite number of singularities. It has
interest in cosmology, as the basis of the model of randomly-placed
power-law halos (Gaite, 2005-A). The locations of these halos may also
be correlated, so that the model actually becomes related to the halo
model proposed here, as we shall see.

\subsection{Correlation moments and multifractal spectrum}
\label{moments}

To analyze the structure of multifractals, it is useful to 
introduce the scale-dependent correlation moments
\begin{eqnarray}
M_q(r) &=& \int dm({\bm x})\; m[B({\bm x},r)]^{q-1}. 
\label{Mq}
\end{eqnarray}
$M_1$ is the total mass, which we normalize to one.  With this
normalization, a multifractal measure can be interpreted as a
probability measure.  $M_2(r)$ is the usual {\em mass-radius} relation
or two-point correlation integral (the integral of the two-point
correlation function). Larger integer values $q = 3, 4, \ldots$ are
also employed in the analysis of the large scale matter distribution,
but they are increasingly difficult to estimate. In addition, since
multifractals are singular non-uniform distributions, integer values
of $q$ are not sufficient and we have to consider the full set of
moments $M_q(r)$ for $-\infty < q < \infty$.  We can then define the
function that gives the scaling behaviour of this full set of moments,
namely,
\begin{equation}
\tau(q) = \lim_{r\ra 0}\frac{\log M_q(r)}{\log r}.
\label{tauq}
\end{equation}
For well-behaved multifractals, this function determines the
multifractal spectrum through a Legendre transform (Falconer, 2003):
\begin{equation}
f(\a) = \mathrm{inf}_q [q\,\a - \tau(q)]\,,
\label{Leg}
\end{equation}
where $\mathrm{inf}_q$ means the infimum with respect to $q$.  If
$\tau(q)$ is differentiable and {\em convex}, the infimum is given by
the vanishing of the derivative with respect to $q$, so
$$\a(q)= \tau'(q).$$ Then, after inverting $\a(q)$, one obtains
\begin{equation}
f(\a) = q(\a)\,\a - \tau[q(\a)],
\label{MFspec}
\end{equation}
which is also convex, namely, $f''(\a) < 0$.

It is useful to define the quantity $D(q) = \tau(q)/(q-1)$, which is
constant for a monofractal, and, in general, decreases with $q$.  It
turns out that the most important values of $q$ are $q = 0,1$.  At $q
= 0$, we have $f(\a_0)= D(0) = -\tau(0)$.  On account that $f'(\a_0)=
q = 0$, the set of singularities with exponent $\a_0$ has the largest
fractal dimension (which often corresponds to the measure's support).
The value $q = 1$ gives the {\em entropy dimension} $D(1)$, for a
value of $\a_1$ such that $\a_1 = f(\a_1) = D(1)$. The
corresponding singularity set defines the {\em measure's concentrate},
that is, the set outside of which the measure vanishes.  Note that
$f'(\a_1) = q(\a_1) = 1$ and convexity of $f(\a)$ imply $f(\a) \leq
\a$ for any $\a$ (with equality at $\a_1$).

For a monofractal, it is sufficient to consider the mass-radius
relation $M_2(r)$ and its exponent $\tau(2) = \a(q) = D(q)$ for all
$q$.  In finite samples from fractals, it is usual to consider the
discrete version of the mass-radius relation, namely, the
number-radius relation $N(r)$, given by the number of particles in a
ball of radius $r$ centered on one particle and averaged over every
particle,

In general, the mass-radius relation $M_2(r)$ has exponent $\tau(2) =
D(2)$. According to Eq.\ (\ref{MFspec}) and the condition $f(\a) \leq
\a$,
$$f(\a_2) = 2 \a_2 - \tau(2) \leq \a_2 \Ra f(\a_2) \leq \a_2 \leq
D(2),$$ where $\a_2 = \a(q=2)$.  We can interpret these relations as
follows.  The multifractal analysis through moments relies on a basic
property: only concentrations with {\em one} exponent contribute to
one $q$-moment, namely, concentrations with the exponent
$\a(q)$. Therefore, to the mass-radius relation only contribute
concentrations with exponent $\a_2$. These concentrations are
distributed in a fractal of dimension $f(\a_2)$.  The exponent $\a_2$
is larger than $f(\a_2)$ due to the presence of other singularities
with different $\a$ in its neighbourhood.  Moreover, since $M_2(r)$ is
obtained by averaging over points, this operation further increases
the exponent.

\subsubsection{Coarse analysis and counts-in-cells}
\label{coarseMF}

The mathematical definition of $q$-moments provided by the integral in
Eq.\ (\ref{Mq}) may not be convenient for calculations. Therefore, in
analogy with the {\em box-counting} method for fractal sets, there is
a method for {\em coarse multifractal analysis} (Falconer, 2003): one
puts an $\d$-mesh of cubes in the total volume (which is itself a large
cube) and considers the mass inside each cube. Then one defines
\begin{equation}
M_q(\d) = \sum_i m_i^{q}\;,
\label{Mq2}
\end{equation}
where the sum is over non-empty cubes.  This method is related to the
coarse-graining method used in cosmology for finite samples under the
name of ``counts-in-cells" (Balian \& Schaeffer, 1989-A): assuming
that all the particles have the same mass, the mass is measured by
counting particles.  In coarse multifractal analysis, the linear cube
size $\d$ defines a scale that one lets go to zero, eventually. In
particular, one lets $\d \ra 0$ to define the function $\tau(q)$
according to Eq.\ (\ref{tauq}) after replacing $r$ with $\d$. From a
mathematical standpoint, the equivalence of both point-centered
results, from Eq.\ (\ref{Mq}), and lattice-based results, from Eq.\
(\ref{Mq2}), is not guaranteed and requires a careful analysis
(Falconer, 2003; Harte, 2001).

Since every set of points in which $\a$ takes a definite value is a
fractal with dimension $f(\a)$, in coarse multifractal analysis, the
number of cells with exponent $\a$ is given by
\begin{equation}
N(\a) \sim \d^{-f(\a)},
\label{Ncell}
\end{equation}
according to the box-counting interpretation of the dimension $f(\a)$.
Let us mention the alternate description provided 
by its number-radius relation
$$ N(\a,r) \sim r^{f(\a)},$$
namely, the average number of cells with exponent $\a$ in a ball of 
radius $r \gg \d$ centered on one cell.
Note that the opposite sign of the exponent in these two numbers is in
accord with their being related to $M_0$ and $M_2$, respectively, and
the equality $\tau(2) = D(0) = -\tau(0)$ for a monofractal.

From relation (\ref{Ncell}), we can obtain the {\em mass function} at
scale $\d$, namely, the number of cells with mass between $m$ and $m +
dm$, which is
$$N(m)\,dm \sim -\d^{-f(\a)} d\a.$$ 
Given that $\a \sim \log m/\log\d$,
\begin{equation}
N(m) \sim \d^{-f(\a)-\a}.
\label{Nm}
\end{equation}

Any measurement of positions of objects (galaxies, etc.)  has to be
made with a given precision, which actually defines the {\em natural}
cell size for the corresponding observation.  Furthermore, a single
object at the given resolution may appear as a compound object at a
higher resolution.  Therefore, coarse multifractal analysis is
naturally adapted to this idea of change in the description of a
distribution in terms of single objects as the resolution increases.
However, let us note that other coarse-graining procedures and, in
general, other methods can be employed for multifractal analysis.  In
particular, let us point out that Halsey et al (1986) define
multifractal analysis by coarse-graining into {\em arbitrary}
(disjoint) pieces $S_1, \ldots, S_N$, such that their diameters
fulfill $\d_i \leq \d$.  Then, the function
$$M_q(\tau,\{S_i\},\d) = \sum_{i=1}^N \frac{m_i^{q}}{{\d_i}^\tau}\;,$$
for small $\d$ (large $N$) and optimized over $\{S_i\}$, is of order
unity only for some value of $\tau$, for given $q$. In the case of
partitioning into an $\d$-mesh of cubes, this definition of $\tau(q)$
coincides with the value given by Eq.\ (\ref{Mq2}).

\subsection{Linear multifractal spectrum: bifractals}
\label{linearMF}

The {\em Legendre spectrum} given by Eq.\ (\ref{Leg}) is only an upper
bound to the real multifractal spectrum (Falconer, 2003).
Nevertheless, for ``reasonable" multifractals they coincide.  However,
there is an interesting situation in which there can be a discrepancy,
as has been noted by Balian and Schaeffer (1989-B). In general, the
Legendre spectrum (\ref{Leg}) must be {\em convex} (from above), but
the real multifractal spectrum needs not be convex. So the Legendre
spectrum associated to a non-convex multifractal spectrum is only its
{\em convex hull}.

For example, let us consider a bifractal, with two exponents $\a_1$
and $\a_2$ (Balian and Schaeffer, 1989-B).  We assume that $\a_1 <
\a_2$.  Although the multifractal spectrum seems to consist of just
two values, namely, $f(\a_1)$ and $f(\a_2)$, the Legendre spectrum is
a full segment with ends at $\left(\a_1,f(\a_1)\right)$ and
$\left(\a_2,f(\a_2)\right)$, that is, a linear spectrum. The
corresponding slope of $\tau(q)$ has a jump from $\a_1$ to $\a_2$ at
$$q_* = \frac{f(\a_2)- f(\a_1)}{\a_2-\a_1}.$$ Now imagine that we 
superimpose on this bifractal another bifractal, either 
$\{\a_1,\a_3\}$ or $\{\a_3,\a_2\}$, such that the
exponent $\a_3$ fulfills $\a_1 < \a_3 < \a_2$ and its
corrsponding dimension fulfills
$$f(\a_3) < \frac{f(\a_2)- f(\a_1)}{\a_2-\a_1}(\a_3-\a_1) + f(\a_1);$$
namely, such that the point $\left(\a_3,f(\a_3)\right)$ is
below the segment with ends at $\left(\a_1,f(\a_1)\right)$ 
and $\left(\a_2,f(\a_2)\right)$.
This second bifractal will be {\em invisible} in both the function 
$\tau(q)$ (which gives the moments) and the Legendre spectrum.

We can consider a linear multifractal spectrum, with $f''(\a) = 0$, as
a limit of general convex spectra, with $f''(\a) < 0$.  To be more
precise, if we are given two exponents $\a_1$ and $\a_2$, and their
respective dimensions $f(\a_1)$ and $f(\a_2)$, convexity implies that
$f(\a)$, for $\a_1 < \a < \a_2$, lies above the segment with ends at
$f(\a_1)$ and $f(\a_2)$.  Assuming that $\tau(q)$ is differentiable,
$\tau'(q)=\a$ is an decreasing function that interpolates from $\a_2$
to $\a_1$ in a range $[q_1,q_2]$. As $q_1 \ra q_2$, $\tau'(q)$ becomes
discontinuous and $\tau(q)$ has an angle, at some value $q_* \in
[q_1,q_2]$. Then $f(\a)$ collapses to the linear spectrum (from
above). In this limit, the multifractal becomes dominated by the two
fractals at the ends of the segment, with dimensions $f(\a_1)$ and
$f(\a_2)$. In other words, this is a {\em bifractal limit} of the
multifractal spectrum.

We can distinguish two cases, according to whether $q_*$ is larger or
smaller than one. In the former case, the measure's concentrate
fulfills $\tau'(1) = \a_2 > \a_1$, so $\a_2 = f(\a_2)$; in the latter
case, the measure's concentrate fulfills $\tau'(1) = \a_1$, so $\a_1 =
f(\a_1)$.  In words, we distinguish whether the measure concentrates
in either the less or the more singular fractal set.  In the former
case, the measure's concentrate and support coincide, whereas in the
latter case, the measure's concentrate corresponds to $\a_1$ and the
measure's support to $\a_2$.  Of course, we can also consider the
borderline case with $q_*=1$.  In this case, $\a = f(\a)$ for $\a_1
\leq \a \leq \a_2$, so the mesasure concentrates uniformly over the
full interval.

Let us mention three interesting examples of linear multifractal
spectrum, corresponding to the three above-mentioned cases.  The first
example is a random distribution of a {\em finite} number of power-law
mass concentrations with exponent $\a < 3$ ({\em one} concentration
suffices). This is a bifractal with exponents $\a$ and 3, and
dimensions $f(\a) = 0$ and $f(3) = 3$, respectively (Gaite,
2005-A). Of course, the measure concentrates in the regular component.
The second example is provided by the one-dimensional adhesion model.
Reasoning about the properties of {\em devil's staircases}, She,
Aurell \& Frisch (1992) and Aurell et al (1997) show that the mass
distribution produced by that model is bifractal: it has singularities
of {\em delta-function} type, that is, with $\a = f(\a) = 0$, and
there appears, in addition, a component with $\a > 1$ and $f(\a) = 1$
(homogeneously distributed, in one dimension).  The extension of these
results to higher dimension is a moot point (Vergassola et al, 1994).
The third example corresponds to the borderline type bifractal: it is
a bifractal formed by the union of two separate monofractals (with
different dimension). In it, the measure is uniformly concentrated,
such that each fractal holds a finite fraction of the total mass.

Let us finally note that any piece of a convex multifractal spectrum
that is not very curved can be approximated by a linear spectrum and,
therefore, by the corresponding bifractal limit. This property will be
useful for the connection with the Press-Schechter formalism (Sect.\
\ref{P-S_sect}).

\subsection{Halos as mass concentrations}
\label{halos-m}

Halo models of large-scale structure assume that the dark matter is in
the form of collapsed halos with definite density profile.  This
profile is established by averaging over angular directions, hence
obtaining a radial profile, whose functional form is essentially the
same for every halo (but with different parameters).  Several forms
have been proposed, the simplest one being a power law $\r(r) \propto
r^{-\b}$, which is also a limit of other more complicated profiles. A
power law halo profile can be related with a power law correlation
function (Peebles, 1974; McClelland \& Silk, 1977; Sheth \& Jain,
1997; Murante et al, 1997).  It is also intrinsically associated with
self-gravity: the Newton-Poisson equations are scale invariant; so we
expect that the mass concentrations produced by them are also scale
invariant.  This argument is valid in the {\em strongly nonlinear}
regime, where the dynamics dominates over the initial conditions. In
particular, the {\em stable clustering} model (Peebles, 1980) supports
this hypothesis (Sheth \& Jain, 1997).

Therefore, we define halos to have a power-law profile, namely, the
{\em singular} profile $\r(r) \propto r^{-\b}$ with $\b = 3 - \a > 0$.
But it is obvious that we cannot assign a definite size to a halo with
this profile.  Indeed, in a multifractal that is scale invariant on
ever decreasing scales, as the definition requires, it is not possible
to assign a size or a mass to any structure, because that would break
scale invariance: any structure must be repeated in every size,
reaching up to the homogeneity scale, which is the only existent
scale.  In consequence, mass concentrations are point-like mass
singularities and they can {\em only} be classified by their
singularity strength, given by the exponent $\a$.  However, a singular
power-law profile cannot be physically realized.  In fact, what is
available is a {\em finite} sample from a multifractal, which is not
scale invariant on ever decreasing scales.

Thus, on small scales, there is a change in the nature of the matter
 distribution, related to the breaking of scale invariance, which
 allows us to assign halos definite size and masses. This change takes
 place in the large scale distribution of matter and, more clearly, in
 cosmological $N$-body simulations, as we shall explain. Of course, on
 scales smaller than the scale of this change, the power-law halo
 profile is altered. Its shape may turn out to fit a different power
 law, thus becoming a combination of two power laws, like in current
 halo profile models (Cooray \& Sheth, 2002).  If we use coarse
 multifractal analysis, the breaking of scale invariance appears below
 a given cell size, which we choose as the size of coarse-grained
 halos.  Then, the number of (coarse-grained) halos is finite.  In
 addition to the full set of halos, with their respective masses, the
 matter distribution is defined by the distribution of their
 positions.  We will treat the distributions of halos in greater
 detail in Sect.\ \ref{fract-dist-halos}.

Let us note that a uniform coarse-graining length is convenient but
not mandatory: in multifractal analysis, one can actually choose
pieces of very different size, as commented in Sect.\
\ref{coarseMF}. In particular, halos can be defined to have different
size and can be given by a different condition. In principle, halos
can be defined
by a condition on the density, as in the usual
definition. The possibility of using various definitions has already
been noted by Valageas, Lacey \& Schaeffer (2000).  We choose a
uniform coarse-graining length, following Vergassola et al (1994). It
is a convenient choice, well adapted to $N$-body simulations. 

\subsection{Voids as mass depletions}
\label{voids}

We define a void as a region with very low or vanishing density, but
we need to make precise what this condition means.  The multifractal
formalism is very useful in this regard.  In this formalism, the
singular concentrations to which we associate halos have a power law
profile $\r(r) \propto r^{-\b}$, with $\b = 3 - \a > 0$. But, if $\a
\geq 3$, there is no singularity.  The value $\a = 3$ ($\b = 0$) is a
reference: a mass concentration with this exponent is not really a
concentration: its density is regular, namely, non-divergent and
non-vanishing. Exponents $\a > 3$ correspond to points of vanishing
density, which we naturally associate with voids.  Furthermore, the
total mass in voids vanishes, since the measure's concentrate, with
$f(\a) = \a$, can only be in regions with $\a \leq 3$ (note that,
necessarily, $f(\a) \leq 3$).

However, any neighbourhood of points in voids, with $\a > 3$, is not
totally empty.  It is possible to have mass around points with
vanishing density. There may or may not be totally empty voids,
according to whether the measure's support occupies the full volume or
not: if it does not, there are regions with non-vanishing volume and
no mass in them (in mathematical terminology, these regions are {\em
open sets}). If the dimension of the support of the measure is smaller
than three, there need to be totally empty voids, and the multifractal
is called {\em lacunar}. In particular, there may be a scaling set of
totally empty voids. This is always the case in one dimension if the
the measure's support has dimension smaller than one, but the geometry
of voids is more complex in higher dimensions (Gaite, 2005-B; Gaite,
2006).

To illustrate the nature of multifractal voids in a simple case, let
us consider again the one-dimensional adhesion model. In this model,
the mass is concentrated in {\em shocks} and their locations form a
{\em dense} set (She, Aurell \& Frisch, 1992; Vergassola et al, 1994).
The adjective ``dense'' is understood with its mathematical meaning: a
set is dense in an interval, say, if in any sub-interval, however
small, there are points of the set; a typical example is the set of
rational numbers between 0 and 1.  In the one-dimensional adhesion
model, while the density at non-shock points vanishes (they have $\a >
1$), these points are surrounded by infinitesimally close small shocks
that make the mass in any neighbourhood of them to be non-vanishing.
Of course, we could make in that distribution a totally empty void by
removing the mass in a sub-interval.

With the preceding definition of void, we can actually consider a
multifractal distribution (in three dimensions) as the union of (i)
halo clusters, that is, clusters of mass concentrations with $\a < 3$,
(ii) voids, that is, mass depletions with $\a > 3$, and (iii) the
boundary between both regions, formed by points with $\a = 3$.
However, we must beware once more of relying on our geometrical
intuition when dealing with multifractals: the set of points with $\a
= 3$ may not be two-dimensional and so it may not constitute a regular
boundary.  Indeed, the dimension of this set is $f(\a=3)$, which can
well be larger than two. We will see in Sect.\ \ref{voids-GIF2} what
this implies for the shape of voids in an example from a
cosmological simulation.

Let us now consider a coarse-grained multifractal.  The
coarse-grained density can be defined to be regular everywhere (with
no singularities) and it only vanishes at a point if the mass vanishes
in a neighbourhood of it.  Therefore, the density only vanishes inside
totally empty voids.  Nevertheless, we can still define voids as
depleted regions comprising coarse-grained exponents $\a_i = \log
m_i/\log \d > 3$ ($\a_i \ra \infty$ inside a totally empty void).
Note that the boundary of voids is then a coarse-grained version of
the above mentioned fractal surface, so it loses small pieces as some
singularities get smeared.

Finally, we notice that the presence of voids in the galaxy
distribution was a principal motivation for the bifractal model of
cosmic structure. Let us quote Balian \& Schaeffer (1989-B): ``The
ocurrence of a bifractal behaviour is due to the coexistence at the
same scales of clusters and voids''. According to our discussion of
the linear multifractal spectra in Sect.\ \ref{linearMF}, we can
deduce that a bifractal in which the voids define one of the two
fractal dimensions must be such that the measure concentrates in the
more singular set and, in addition, such that the more regular set has
$\a_2 > 3$ and is associated with the voids.  An interesting
one-dimensional example is the adhesion model, according to Sect.\
\ref{linearMF}.  However, ``coexistence at the same scales of clusters
and voids'' takes place in most multifractals, not only in bifractals.

\section{The density probability distribution function}
\label{PDF}

In coarse multifractal analysis, moments are defined by Eq.\
(\ref{Mq2}).  There is a related way to define statistical
moments, namely, by using the density probability distribution
function, and this definition has been applied to multifractals
(Borgani, 1993; Valageas, 1999). If $P(\r)$ is the probability of
having density $\r$, then, its moments are
\begin{equation}
\mu_q = \langle \r^q \rangle = \int_0^\infty d\r \,\r^q P(\r).
\label{Mq3}
\end{equation}
Since in multifractals the density does not exist at singular points,
$\r$ in this definition must be also understood as a {\em
coarse-grained density}. Besides, this definition implies that $\mu_0
= 1$, assuming $P(\r)$ normalized. In contrast, $M_0 \neq 1$, in
general, and it is related with the fractal dimension of the measure's
support.  In coarse multifractal analysis, the precise relationship
between both definitions of $q$-moments is
$$\mu_q(\d) = \langle \r^q \rangle = \frac{\sum_i \r_i^{q}}{\sum_i 1}
= \d^{-3q}\frac{M_q(\d)}{M_0(\d)} \sim \d^{\tau(q) - \tau(0) -3q}\,$$
(where the sums are over non-empty cells).  For example, $\mu_1(\d)$
is the average density (within the measure's support) and scales with
exponent ${- \tau(0) -3} = D(0) - 3 \leq 0$. If the measure's support
has dimension 3, then the average density is $\d$-independent. In this
case, it is natural to normalize the density field by dividing it by
the average density, so that $\mu_1 = 1$ (Valageas, 1999).  If the
measure's support is fractal, then the average density diverges as $\d
\ra 0$.

Regarding higher moments $\mu_q(\d), \;q >1,$ we have
$$
\frac{\mu_q(\d)}{\langle \r \rangle^q} \sim \d^{(q-1)[D(q) - D(0)]}.
$$ 
Given that $D(q)$ is a non-increasing function, this quotient is
finite, in the limit $\d \ra 0$, only for a {\em uniform}
multifractal, namely, a regular distribution or a monofractal (we say
that a monofractal is uniform because the exponent $\a$ is constant).
For a generic multifractal, that quotient diverges.

One can derive other relations between moments.  For example,
\begin{equation}
\frac{M_q(\d)}{M_2(\d)^{q-1}} \sim \d^{(q-1)[D(q) - D(2)]}\,,
\label{hierarchy}
\end{equation}
which diverges for $q > 2$, except in uniform multifractals, namely,
monofractals (including regular distributions, with $\a=3$).  Thus,
this latter relation is useful in the context of {\em hierarchical
models}: a scale-independent quotient $M_q/M_2^{q-1}$ implies $D(q) =
D(2)$.  In other words, ``hierarchical correlation functions imply
monofractality of the distribution'' (Borgani, 1993).  An analogous
relation holds for $\mu_q(\d)$ if $D(0)= 3$.

The probability distribution function (PDF) of the coarse-grained
density, $P_\d(\r)$, is directly related to the mass function and,
therefore, to the coarse multifractal spectrum (Sect.\
\ref{coarseMF}): the number of cells with density $\r$ is the
box-counting approximation to the fractal dimension of the set of mass
concentrations with exponent $\a = \log m/\log \d = \log \r/\log \d +
3$.  The general statement of this relation is best made in the
context of probability theory, namely, of the theory of {\em large
deviations}.  The relevant concepts in this theory appear naturally
when we consider the generalization of the central limit theorem to
nonlinear distributions and, hence, we regard the lognormal
distribution as a general nonlinear PDF.

\subsection{Connection with the lognormal model}
\label{lognormal}

As a simple extension of the Gaussian density field in the linear
regime, Coles \& Jones (1991) proposed that a lognormal field model
would be suitable for the nonlinear regime, especially, in the {\em
weakly} nonlinear regime. We shall see that a lognormal density PDF is
natural in connection with multifractals.

From a purely statistical point of view, the Gaussian distribution is
associated with the central limit theorem and one expects that it
appear in many situations where the {\em sum} of a large number of
independent random variables is involved. A nonlinear analogue of this
theorem is provided by the {\em multiplication} of independent random
variables, because multiplication of these variables is equivalent to
addition of their logarithms. Hence, the lognormal distribution arises
as the limit distribution in multiplicative processes. This
formulation of the lognormal distribution corresponds to the {\em
large-deviation} formulation of multifractals (Frisch, 1995; Sornette,
2000; Harte, 2001), which is particularly adapted to random cascade
processes but has a broader scope. We proceed to introducing a few
relevant notions, leaving aside innecessary details.

We may recall that the central limit theorem considers the limit
distribution of the sum of $n$ independent random variables with
finite variances (and zero mean). It states that this limit
distribution is Gaussian, with variance given by the sum of the
variances. In the particular case of $n$ independent identically
distributed variables, their sum divided by $\sqrt{n}$ has a Gaussian
distribution (with the common variance).  However, if we want to take
into account unlikely events, that is deviations of magnitude larger
than $\sqrt{n}$, we must consider higher moments than the variance.
In particular, if we call $X$ the sum of the $n$ variables, as $n \ra
\infty$,
$$P[X \sim n\,x] \sim e^{n\,S(x)}.$$ In this formula, $x$ is an
``intensive'' variable, that is, finite in the limit $n \ra \infty$,
and $S(x)$ is an entropy function called the Cram\'er function. This
function can be obtained from the full (common) distribution function
of the variables (Frisch, 1995; Sornette, 2000). Its second order
expansion at its maximum (assumed unique) gives the Gaussian
approximation that constitutes the central limit theorem.

In multiplicative processes, such as random cascades or fragmentation
processes, we also have that a good approximation is given by the two
basic moments (mean and variance) of the logarithm of the relevant
random variable. Therefore, the lognormal distribution appears as the
basic approximation to the distribution of mass concentrations in a
multifractal. This has already been noted by Jones, Coles \&
Mart\'{\i}nez (1992). The large-deviation formalism becomes necessary
if we need to consider mass concentrations that appear rarely, because
of their small or large exponent $\a$ with respect to $\a_0$, the
value that maximizes $f(\a)$ (assumed unique).  In fragmentation
cascades, the number $n$ in the hierarchy gives the scale, for
example, $\d_n = 2^{-n}$ for a binary division. The sum variable $X$
represents minus the logarithm of the mass, such that the ``intensive"
variable is now the exponent $\a = \log m/\log\d_n$.  Therefore,
$$P[-\log_2 m \sim n\,\a] \sim e^{n\,S(\a)},$$ and we are to identify
$S(\a)$ with the multifractal spectrum $f(\a)$; namely, $S(\a) =
f(\a)\,\ln 2$.  Thus the large-deviation condition is equivalent to
focusing on mass concentrations with exponents $\a$ such that their
supporting regions have small dimension $f(\a)$ compared with its
largest value.  In the ``small-deviation'' regime, we can use the
second order expansion (parabolic approximation) of $f(\a)$ at $\a_0$,
the value that maximizes it (assumed unique).  This value corresponds
to the measure's support.  The ``small-deviation'' regime is indeed
the weakly nonlinear regime, and it gives a lognormal PDF.

Note that Jones, Coles \& Mart\'{\i}nez (1992) chose to take a
parabolic approximation to $f(\a)$ that is not its second order
expansion but instead makes it tangent to the diagonal, like the full
$f(\a)$. This choice does not seem related to the theory of large
deviations.  However, a large-deviation expansion around the measure's
concentrate (the point of tangency to the diagonal) is possible as
well. Let us consider the mass moment $M_1$ and its associated mass
PDF
$$m \,P[-\log_2 m \sim n\,\a] \sim m\,e^{n\,S(\a)} \sim
e^{n\,[f(\a)-\a]\,\ln 2}.$$ 
In the limit $n \ra \infty$, this density vanishes unless $f(\a) =
\a$, that is, unless in the measure's concentrate. In other words,
there is no mass outside the concentrate of the measure (mass),
according with its name.  Nevertheless, we can consider again a
``small-deviation'' regime given by the second order expansion around
the measure's concentrate, such that it gives a lognormal mass
PDF. Note that this lognormal mass PDF is generally different from the
previous lognormal PDF.

We must mention that the lognormal approximation to multifractals
assumes that the multifractal spectrum $f(\a)$ is well behaved,
namely, that it has second derivative.  Of course, this condition of
differentiability may not hold in some cases.  In particular, a
bifractal spectrum does not have well-defined second derivative at its
maximum.

It is well known that the lognormal distribution is not determined by
its moments $\mu_n, \; n = 1,2, \ldots $, that is, there are other
distributions with the same set of moments. In physical terms, the
reason for this non-uniqueness can be ascribed to the fact that
high-order moments are dominated by high-density regions, so that the
full set of moments contains insufficient information about very
low-density regions or {\em voids} (Coles \& Jones, 1991). This was
the motivation for considering the additional information on voids,
namely, the void probability function, as a necessary ingredient.  In
fact, the combination of separate information on clusters and voids,
respectively, and the assumption of scale invariance led Balian \&
Schaeffer (1989-B) to propose a {\em bifractal} model of large scale
structure. From the preceding discussion about the lognormal model and
its relation with multifractals, we can deduce that Coles \& Jones's
(1991) and Balian \& Schaeffer's (1989) arguments are related and that
we do not need to restrict ourselves to bifractals. Indeed, we have
already pointed out that the singular nature of multifractals requires
us to use the full set of moments $M_q(r)$, for $-\infty < q <
\infty$. In particular, negative values $q < 0$ are typically
associated with voids.

We have explained the connection between multifractal and the
lognormal model of Coles \& Jones (1991) in regard to the density PDF.
Further connection in regard to $n$-point correlation functions, if
possible, seems to require additional assumptions. We note, in
particular, that there is {\em no scale invariance} in the lognormal
model as a field theory, in spite of the arguments presented by Coles
\& Jones (1991) regarding a possible connection with multifractality.

\subsection{Connection with the Press-Schechter mass function}
\label{P-S_sect}

The usual mass function in halo models is the one given by the
Press-Schechter spherical collapse formalism (Mo \& White, 1996; Sheth
\& Jain, 1997).  This formalism relies on an initial Gaussian PDF.
Regarding the relation that we have established in multifractals
between the mass function of halos and the coarse density PDF (for $\a
< 3$), we shall explore if there is a connection with the
Press-Schechter mass function.  Let us briefly recall the relevant
elements of this formalism.

Let us consider the coarse density and its PDF. For early times or
large values of the coarse-graining length $\d$, we can consider the
density constant ($\r = 1$, if the total volume and mass are
normalized) except for small fluctuations given by a Gaussian PDF.
This Gaussian PDF has a scale-dependent variance $\s^2(\d) \ll 1$.  To
obtain the mass function of collapsed objects, one needs a criterion
to decide which regions of size $\d$ will collapse. The simplest
criterion is that every region in which the density exceeds some
threshold $\r_c$ must collapse. Given that the PDF is Gaussian, the
cumulative distribution function is
\begin{eqnarray}
P(\r > \r_c) &=& \frac{1}{\sqrt{2\pi}\,\s} 
\int_{\r_c}^\infty \exp\left(-\frac{(\r-1)^2}{2\,\s^2}\right)\,d\r
\nonumber\\
&=& 
\frac{1}{2}\left[1-\mathrm{erf}\!
\left(\frac{\r_c-1}{\sqrt{2}\,\s}\right)\right].
\label{P-S_eq}
\end{eqnarray}
The mass of regions of size $\d$ is $m \sim \d^3$.  Therefore, we can
express $\s^2(\d)$ in Eq.\ (\ref{P-S_eq}) in terms of the mass of
collapsed objects, provided that this equation gives the fraction of
regions of mass greater than $m$ that collapse for a given threshold
$\r_c$.  We may call this fraction $F_>(m)$. It is a function of the
ratio $(\r_c-1)/\s$, so that raising the threshold $\r_c$ at fixed
$\s$ is equivalent to diminishing $\s(\d)$ at fixed $\r_c$ and making
density peaks smaller.  $\s(\d)$ must be a decreasing function of $\d$
and, therefore, of $m$. Thus, collapse of large masses is suppressed.

The mass function is 
\begin{eqnarray}
N(m)  
= - \frac{dF_>(m)}{dm} \,\frac{1}{m} = 
-\frac{1}{m}\,\frac{dF_>(m)}{d\s} \,\frac{d\s}{dm} 
\nonumber\\
= 
\frac{1}{\sqrt{2\pi}} 
\frac{1}{m}\,\frac{d\s}{dm} 
\frac{\r_c-1}{\s^2}
\exp\left(-\frac{(\r_c-1)^2}{2\,\s^2}\right),
\end{eqnarray}
where we have introduced the factor $m^{-1} = \d^{-3}$ which
represents the total number of collapsed objects. We assume an initial
power-law spectrum of Gaussian fluctuations, namely, $\s^2(\d) \propto
\d^{-n-3}$ ($n > -3$ is the spectral index in Fourier space). Then,
the mass function is also a power law, but with an exponential decay
for large masses; namely,
\begin{equation}
N(m) \propto \left(\frac{m}{m_*}\right)^{n/6 - 3/2}
\exp\left[-\left(\frac{m}{m_*}\right)^{n/3 + 1}\right],
\label{P-S_MF}
\end{equation}
where $m_*$ stands for the large-mass cutoff.

We must now consider carefully the limitations of the Press-Schechter
formalism.  For this, me must distinguish the one-dimensional case
from the others, including the physical three-dimensional case.  The
essential assumptions are the validity of an initial Gaussian PDF and
of the spherical collapse model. A Gaussian PDF is valid if $\d$ is
large, namely, $\s^2(\d) \ll 1$. The spherical collapse model amounts
to assuming that every overdensity above the threshold gives rise to a
symmetric collapsed object. This is a reasonable assumption in one
dimension, but it is questionable in higher dimensions.  In fact, only
a small percentage of the initial overdensities undergo nearly
spherical collapse, and they are precisely the largest overdensities
(Audit, Teyssier \& Alimi, 1997; Lee \& Shandarin, 1998).  This is
particularly clear in the Zeldovich approximation.  Thus, the
Press-Schechter formalism can only apply, in three dimensions, to the
formation of the strongest concentrations (as already noted by
Vergassola et al (1994) in their comparison of the Press-Schechter
formalism with the adhesion model).  The Press-Schechter formalism has
been extended to ellipsoidal collapse, with the consequent improvement
(Sheth \& Tormen, 1999).  However, this extension still excludes
irregular collapses, which are common for weak concentrations.

In conclusion, the Press-Schechter formalism produces a power-law mass
function with a large-mass cutoff, but it can only be applied to massive
halos.  Considering this formalism as an approximation to the full
nonlinear halo mass function based on the linear theory, we would like
to explore its relationship with a lognormal PDF, which is valid
in the weakly nonlinear regime.  Then the question is if relatively
strong mass concentrations can in fact have a power-law mass function
(neglecting the exponential cutoff) that is consistent with an overall
lognormal-like distribution.

To answer this question, let us mention firstly that it is known that
the lognormal distribution can be mistaken for a power law over a
relatively large interval (Sornette, 2000, page 80).%
\footnote{This fact has also been noted in the astrophysical
literature, in a small-scale process: star formation from cloud
fragmentation. The theory of Zinnecker (1984) leads to a lognormal
mass function, as mentioned by Coles \& Jones (1991). Zinnecker (1984)
compares the lognormal mass function with Salpeter's power law.}  As a
parabolic approximation to a multifractal spectrum, the lognormal
distribution derives from the second order expansion around its
maximum, representing the measure's support.  But the measure's
concentrate is best represented by the second order (parabolic)
approximation around the point $f(\a) = \a$, as explained in Sect.\
\ref{lognormal}. This second parabola can actually be very close to
the first one. In any event, an even simpler approximation is the
first order (linear) approximation around the point $f(\a) = \a$.

In fact, a poor sampling of a multifractal will mainly reflect the
measure's concentrate. Let $\a_1 = \a(1)$; given that $\a_1 =
f(\a_1)$ and $f'(\a_1) = 1$, the linear approximation around the
measure's concentrate has unit slope, such that it becomes an even
bifractal (not loaded toward either the more or the less singular end,
as explained in Sect.\ \ref{linearMF}).  In contrast, note that the
power-law part of the Press-Schechter mass function, Eq.\
(\ref{P-S_MF}), produces a divergence in its integral over $m$ in the
limit $m \ra \infty$. Of course, this divergence is killed by the
exponential factor.  Nevertheless, it indicates that {\em the mass
concentrates on massive halos}.  In this regard, the Press-Schechter
mass function is analogous to the mass function of the one-dimensional
adhesion model (according to our discussion in Sect.\
\ref{linearMF}). This is not surprising: the one-dimensional adhesion
model is the simplest mathematical description of the notion of
collapse of overdensities, realizing a collapse into delta-type
singularities (Vergassola et al, 1994).

\section{The fractal distribution of halos}
\label{fract-dist-halos}

The realization of scale invariance in a multifractal is characterized
by the multifractal spectrum.  In Sect.\ \ref{coarseMF}, we have shown
its relation to the mass function, which is related in turn to the PDF
$P_\d(\r)$. This PDF can be represented by the statistical moments
$\mu_q(\d)$ (Sect.\ \ref{PDF}).  However, in cosmology, it is normal
to measure correlations of a given class of objects rather than
moments of the total dark matter distribution. More precisely, one
measures positions of luminous objects, namely, galaxies, and computes
their correlation functions to establish the statistical properties of
the distribution, which may include scale invariance, etc. One can
associate galaxies with dark matter halos and, therefore, deduce
statistical properties of the distribution of these halos; and
viceversa.  This is why the distribution of halos has special
interest.

If the full dark matter distribution is scale invariant, then the
correlation functions of the distribution of halos must be power laws.
More precisely, in a multifractal distribution, every population
formed by similar halos is a monofractal, although different populations
have different dimensions (Gaite, 2005-A). We can describe this
difference between populations as a kind of {\em bias}, albeit of {\em
non-linear} type.

Actually, the very definition of multifractal implies its
interpretation as a set of fractal distributions of mass
concentrations (and depletions), classified according to the different
values of the exponent $\a$: a set of concentrations with exponent
$\a$ forms a monofractal of dimension $f(\a)$. However, for a given
exponent $\a_*$, the corresponding set of concentrations generally has
dimension $f(\a_*) < \a_*$, whereas a monofractal fulfills the
equality $\a_* = f(\a_*)$. This happens because in the neighbourhood
of the concentrations with $\a = f(\a_*)$ are present other
concentrations with $\a \neq f(\a_*)$, the mass of which raises the
singularity exponent from $f(\a_*)$ to $\a_*$.

In the coarse-grained formulation with fixed halo size $\d_h$, we have
a finite number of individual halos which we can classify by their
masses. Then, each class defines a population and we can compute its
number-radius relation, $N(\a,r), \; \a \sim \log m/\log \d_h$ (Sect.\
\ref{coarseMF}).  The fractal dimension obtained from each
number-radius relation is the only quantity necessary to determine the
corresponding monofractal scaling. Thus, the full set of dimensions
obtained in this way constitutes an approximation to the multifractal
spectrum (in the range $\a < 3$).

We have defined halos by relying on coarse multifractal analysis with
a given cell size. In contrast, the measure $N(\a,r)$ is
point-centered, like the mass-radius relation, given by Eq.\
(\ref{Mq}) with $q=2$. However, this combination of lattice and
point-centered methods is consistent.  Nevertheless, one can also use
the coarse-grained probability $P_\d(\r)$, providing information on
the number of boxes of size $\d$ with mass $m \sim \d^\a$ and, hence,
on the multifractal spectrum (as explained in Sect.\ \ref{lognormal}).
Using several values of $\d$, we would reproduce the {\em histogram
method} of calculating the multifractal spectrum (Falconer, 2003),
which indeed consists of making histograms of the $\a_i$ for several
$\d$. This method is, in general, less suitable than the method of
moments (which uses the $M_q(\d)$ of Eq.\ (\ref{Mq2})).  In fact, ee
have found most convenient the procedure that consists of (i) the
definition of halo populations at given $\d_h$, and (ii) the
calculation of their respective number-radius relations $N(\a,r)$.
The explanation lies on the effect on correslations of a limited
scaling range, as we show next.

\subsection{The scaling of correlation moments}
\label{moment-scaling}

In general, the mass distribution can be described by either the
coarse-grained probability $P_\d(\r)$ or the moments $M_q(r)$ (or
$\mu_q(\d)$).  In particular, its multifractal properties can be
described by either $f(\a)$ or $\tau(q)$, which are related by the
Legendre transform (Sect.\ \ref{moments}) and provide identical
information, {\em in principle}.  The formulation of a multifractal as
a fractal distribution of halos emphasizes the scale invariance of
each halo population; in particular, the number-radius relation
$N(\a,r)$ expresses the scaling properties of each halo population as
a monofractal, such that the full set of exponents of all populations
constitutes an approximation to the multifractal spectrum (in the
range $\a < 3$). Therefore, the full set of exponents also provides an
approximation to $\tau(q)$, in the corresponding range of $q$.

The full mass-radius relation $M_2(r)$ is directly related with the
reduced two point-correlation $\xi(r)$, such that scale invariance
implies that they both are power laws.  Unfortunately, the analysis of
the two-point correlation function in cosmological $N$-body
simulations by the Virgo Consortium has not shown any convincing
scaling (Jenkins et al, 1998).  We do not consider this negative
result as ruling out scale invariance.  In fact, we can understand it
as an effect of a limited scaling range in a distribution that is
actually multifractal, rather than monofractal.  Let us explain how
that effect arises.

The equivalence between $f(\a)$ and $\tau(q)$ relies on the one-to-one
dependence $\a(q)$.  However, this one-to-one dependence only holds in
a mathematical multifractal (in the limit of ever decreasing scales).
In the coarse-grained formulation, we can easily deduce that the
average over populations implied by the definition of $M_q(r)$, Eq.\
(\ref{Mq}), will spoil the scaling of {\em each} particular population
(with a given exponent $\a$). In other words, in the coarse-grained
formulation, every halo population contributes to some extent to a
$q$-moment (in particular, to $M_2(r)$), rather than only the
population with $\a(q)$. For this reason, to realize scale invariance,
it is preferable to analyse the fractal distributions of halos {\em
independently}, rather than the $q$-moments of the total distribution.

\subsection{The transition to homogeneity}
\label{select}

So far, we have used the definitions of scaling exponents, for
example, Eq.\ (\ref{a}), as impliying naive scaling relations, such as
$m \sim r^\a$.  Of course, this relation holds in a limited range of
scales (namely, of $\log r$).  In addition to the restriction of
scaling on small scales, already studied, there has to be a transition
to homogeneity on large scales.  Over the scale of this transition,
say $r_0$, the relation $m \sim r^\a$ undergoes a crossover to $m \sim
r^3$ (homogeneity).  Therefore, we can define $r_0$ by the equation
\begin{equation}
m = \frac{4\pi\bar\r}{3}\,r_0^3 \left(\frac{r}{r_0}\right)^\a, 
\; r \ll r_0\,,
\label{m-r0}
\end{equation}
where $\bar\r$ is the large-scale mean density.
Analogously, the halo number-radius relation $N(\a,r) \sim r^{f(\a)}$
(Sect.\ \ref{coarseMF}) can also be written as
\begin{equation}
N(\a,r) = \frac{4\pi\,\bar{n}(\a)}{3}\, r_0^3
\left(\frac{r}{r_0}\right)^{f(\a)}, \; r \ll r_0,
\label{N-r0}
\end{equation}
where $\bar{n}(\a)$ is the large-scale mean density of halos of
exponent $\a$.

The procedure to determine $r_0$ from $N(\a,r)$ consists of an
analysis of the crossover to homogeneity in its log-log graph.  It
must be done for several values of $\a$, to check that the results
coincide.  Actually, the extreme cases are most useful; that is to
say, we may take, on the one hand, (a) the smallest value of $\a$ (the
largest halo mass) such that there is a sufficient number of halos to
compute $N(\a,r)$, and, on the other hand, (b) the largest value of
$\a$ (the smallest halo mass) such that its fractal dimension is
sufficiently smaller than three as to let us clearly perceive the
crossover to homogeneity.  In Sect.\ \ref{fracr-dist-h}, we shall see
how to carry out this procedure in an example.

\section{Halos and voids in cosmological simulations}

In $N$-body cosmological simulations, the nature of the distribution
changes on small scales because there is (i) a minimal mass,
corresponding to one particle, and (ii) a softening of the
gravitational force.  The minimal mass defines the discretization
scale, namely, the linear size of the volume per particle. The larger
of the discretization scale and the softening scale, usually, the
discretization scale, is the natural (intrinsic) coarse-graining scale
of the distribution.

Since all the particles have the same mass, the mass is measured by
counting particles.  We have already defined {\em coarse multifractal
analysis} and mentioned its relation with the method of
``counts-in-cells" (Balian \& Schaeffer, 1989-A), whereby one puts a
$\d$-mesh of cubes in the total volume (which is itself a large cube)
and measures the mass inside each cube (cell) by counting particles.
In this method, initially and as long as the evolution is linear,
cells of the size of the discretization scale contain one particle per
cell.  Therefore, halos only arise in the nonlinear stage, as they
concentrate particles from other regions that become {\em voids}.

We define mass concentrations (or depletions) at scale $\d$ and
measure their exponents by the equation $\a_i = \log m_i/\log \d$,
where $\d$ is the cell size. Standard coarse multifractal analysis
proceeds to either (i) make histograms of the $\a_i$ for several $\d$
(the {\em histogram method}) or (ii) calculate the moment sums
$M_q(\d)$ in Eq.\ (\ref{Mq2}) (Falconer, 2003). Either way leads to an
estimation of the multifractal spectrum.  However, here we shall
combine the ``counts-in-cells" method with point-centered methods,
according to Sect.\ \ref{fract-dist-halos}.

We have applied our numerical methods to several $N$-body cosmological
simulations. We present the results of the analysis of the redshift
$z=0$ positions of the $\L$CDM GIF2 simulation (by the Virgo
Consortium), with $400^3$ particles in a volume of (110 $h^{-1}$
Mpc)$^3$. The force-softening length is $\e = 7 \,h^{-1}$ kpc.  This
simulation is described by Gao et al (2004), who also use it for halo
analyses, though they employ the usual FoF definition.  Naturally, the
discretization length is $400^{-1}$ (in box-length units) and is
considerably larger than the softening length.  We determine from it
the appropriate halo coarse-graining scale: since we operate with
powers of two, we take as halo size $\d_h = 256^{-1}$. Actually, our
results are not very sensitive to the precise value of $\d_h$.

\subsection{Counts in cells and halos}
\label{counts}

\begin{figure}
\includegraphics[width=7.5cm]{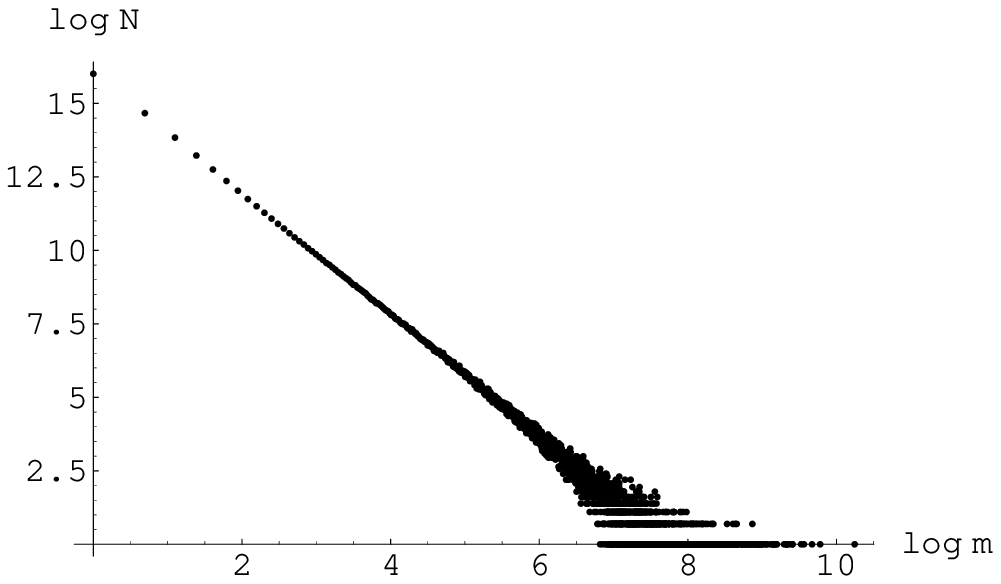}\\
\includegraphics[width=7.5cm]{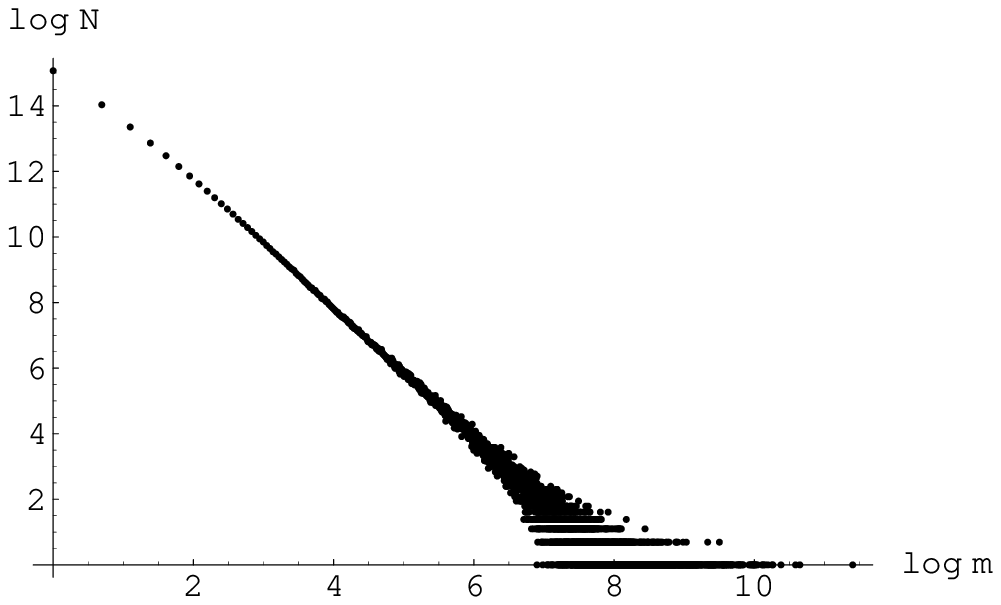}\\
\includegraphics[width=7.5cm]{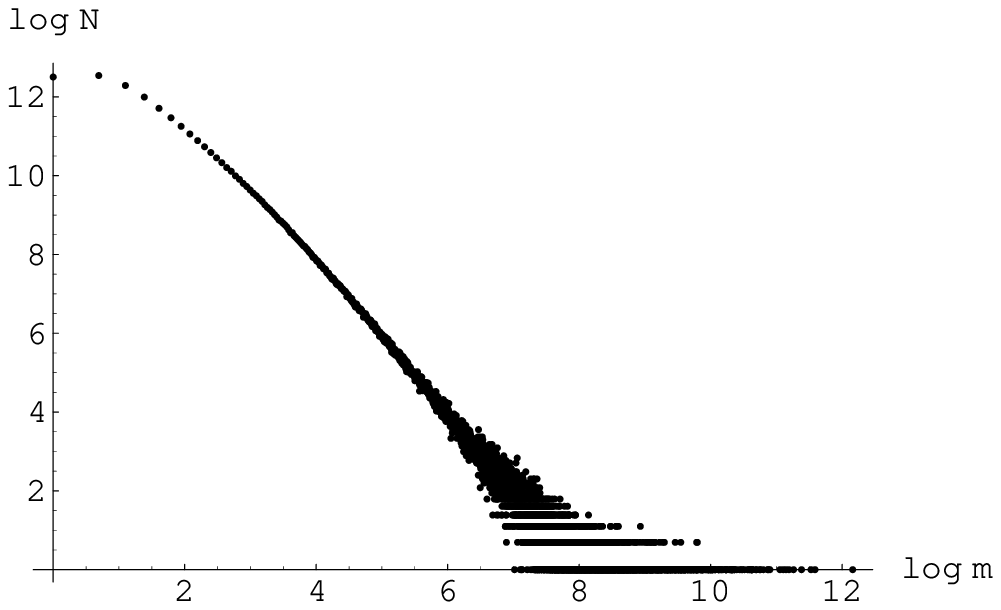}
\caption{Log-log plots of number of halos $N$ versus 
their mass $m$ (number of particles)
at coarse-graining scales $512^{-1}$ (top), $256^{-1}$ (middle), and 
$128^{-1}$ (bottom).  
Both the top and middle plots are linear, but the latter, 
corresponding to lower resolution, already shows a slight 
decay in the number of light halos. The bottom plot clearly shows a 
maximum for $m=2$.
}
\label{P-S}
\end{figure}

In the method of ``counts-in-cells", the qualitative effect of
discreteness dependes on the cell size, as we study next.  If this
size is too small, namely, much smaller than the force-softening length,
most cells are empty an the non-empty ones have
only one particle, so this is the regime totally dominated by
the discretization and the gravitational softening.  
As the size grows, the ratio of non-empty cells grows
as well, and some of them begin to have more than one particle.  In
fact, as the cell size grows, soon a few cells become massive (while
the great majority are still empty), in such way that a pattern
appears: the number of cells with given mass $N(m)$ becomes a power
law.  This is shown in Fig.\ \ref{P-S}.%
\footnote{The power-law dependence of $N(m)$ implies, in particular,
that the cumulative cell number $N_>(m)$ (the number of cells with
mass larger than $m$) also follows a power law. Since $N_>(m)$ is the
rank, we have that the rank-ordering of cells follows a power law,
that is, it satisfies Zipf's law (Zipf, 1949; Sornette, 2000). This
property also applies to halos, according to our definition. The
connection of the Zipf law for voids with fractality has been studied
by Gaite \& Manrubia (2002) and Gaite (2005-B, 2006).}  This regime is
still dominated by the discretization.

As the cell size grows larger, at some point, the most numerous cells
are no longer the ones with $m=1$ (Fig.\ \ref{P-S}).  Further increase
of the coarse-graining scale eventually leads us to a lognormal-like
distribution.  We note that this last transition, namely, from the
power-law regime to the lognormal-like regime, takes place at a
coarse-graining length aproximately equal to $\d_h = 256^{-1}$ (Fig.\
\ref{P-S}).  Thus, this is the scale at which we have maximal richness
and variety of halos, that is, they span the maximal mass-scale
range. On the contrary, at this scale, voids are hardly sampled.

Within the power-law regime of $N(m)$, the power law holds very
clearly. For example, in the top plot in Fig.\ \ref{P-S}, in the
horizontal axis (log m), the scaling holds from 0 to 7, that is, from
$m = 1$ to $m = e^7 = 1097$, namely, three decades (and larger range
on the vertical axis).  We have performed least-squares fits of $\log
N$ versus $\log m$ to find the power-law exponents.  To minimize the
standard error of the fits, it is convenient to reduce the scaling
range.  The fit corresponding to $\d = 256^{-1}$ in the range $m \in
[10,100]$ yields $-1.996 \pm 0.004$; the fit corresponding to $\d =
512^{-1}$ yields $-2.022 \pm 0.004$.  We conclude that $N(m) \sim
m^{-2}$.  This power law amounts to a linear multifractal spectrum,
according to Eq.\ (\ref{Nm}). Moreover, it is consistent with $\a =
\a_1$, corresponding to the measure's concentrate (such that $\a_1 =
f(\a_1)$ and $f'(\a_1) = 1$).

The counts-in-cells method of computing $N(m)$ for various
coarse-graining lengths would allow us to estimate the multifractal
spectrum (according to the histogram method). However, the
counts-in-cells formalism is especially suited to compute the moment
sums, which provide a more accurate estimation.  In fact, it is best
to replace in Eq.\ (\ref{Mq2}) for $q$-moments the sum over cells with a
sum over number of particles:
\begin{equation}
M_q(\d) = \sum_{m=1}^{\infty} N(m)\,m^{q}\,.
\label{Mq4}
\end{equation}
This formula is a discrete version of Eq.\ (\ref{Mq3}).
In a cosmological simulation, the upper limit to $m$ in the sum is, of
course, the total number of particles $N_{\mathrm{tot}}$ ($400^3$ in
the GIF2 simulation). In addition, we have to introduce a
normalization factor to have $M_1 = 1$; namely, we have to divide $m$
by $N_{\mathrm{tot}}$.

We can deduce from Eq.\ (\ref{Mq4}) how halos of different mass
contribute to $q$-moments. Since we have found that the halo mass
function $N(m) \sim m^{-2}$, $q$-moment are given by sums of
$q-2$-powers. In these sums, the contribution of each term increases
or decreases with $m$ according to whether $q$ is larger or smaller
than 2.  In particular, every term of $M_2(\d)$ contributes equally.
This blatantly contradicts the assumption that only contribute to
$M_2(\d)$ singularities with a given exponent ($\a_2$) and, therefore,
a given $m$. 

\begin{figure}
\includegraphics[width=8cm]{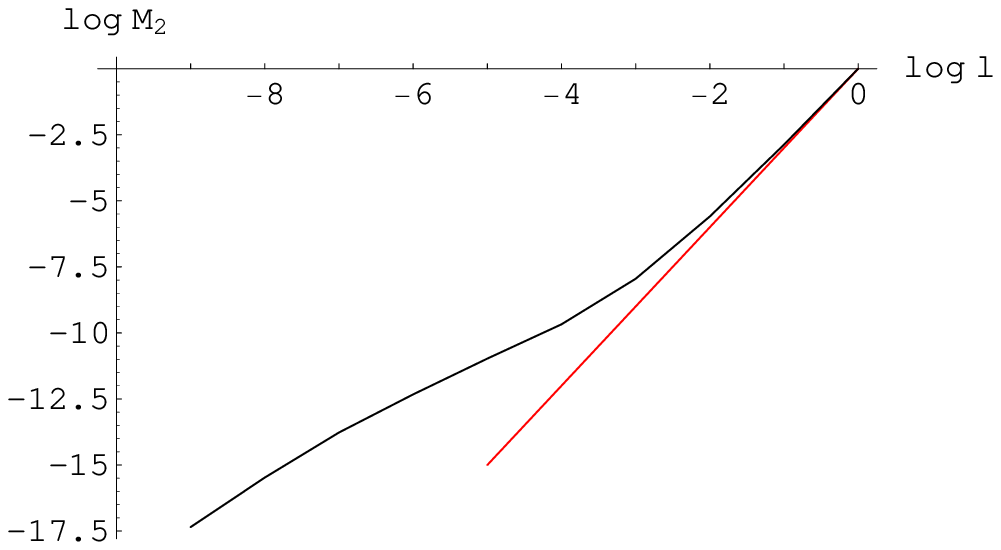}
\caption{Log-log plot of $M_2(\d)$ (logarithms are to base $2$ and the
total size is normalized to unity).  Note the reduced, hardly 
noticeable scaling range (the
straight line corresponds to homogeneity).}
\label{fig_M2}
\end{figure}

The assumption that only one $\a$ contributes, namely,
$\a(q)$, is crucial for the scaling of moments, as commented in Sect.\
\ref{moment-scaling}.  We display in Fig.\ \ref{fig_M2} the log-log
plot of $M_2(\d)$, for $-9 \leq \log_2 \d \leq -1$, to show the
absence of a convincing scaling range.  Indeed, to reach full
multifractal scaling, we need to be in the lognormal-like
regime of the halo mass function $N(m)$. But this regime appears for
cell sizes larger than $\d_h$ and up to the transition to homogeneity,
comprising a scale range that is insufficient for the scaling of
$q$-moments.  Therefore, it is natural that no scaling range of the
two-point correlation function was found in the analyses of $N$-body
simulations by the Virgo Consortium (Jenkins et al, 1998).  In
contrast, scaling is found in the distributions of halos, as we show
next.

\subsection{Fractal distributions of halos}
\label{fracr-dist-h}

The description of multifractal large-scale structure in terms of
fractal distributions of halos has been explained in Sect.\
\ref{fract-dist-halos}. In brief, given the halo size, halos are
classified by their masses (or by their exponents, which is
equivalent). Then we can directly test if each class of halos is a fractal
set by calculating its two-point correlation function, or, rather, its
number-radius relation.

The study of fractal distributions of equal-mass halos serves two
purposes: first, to provide us with a proof of multifractality and,
second, to obtain the homogeneity scale.  According to Sect.\
\ref{select}, to obtain the homogeneity scale, it is convenient to
select two halo populations with extreme values of the halo mass 
(but the light halo population must not be too homogeneous).
Actually, we need a range of masses in each population to have a
sufficient number of halos to compute the number radius relation.  We
choose two ranges, with 750 to 1000 particles and with 100 to 150
particles, defining two populations consisting of 2508 and 25556
members, respectively (see Fig.\ \ref{P-S}, middle plot).

\begin{figure}
\begin{minipage}{7.5cm}
\includegraphics[width=7cm]{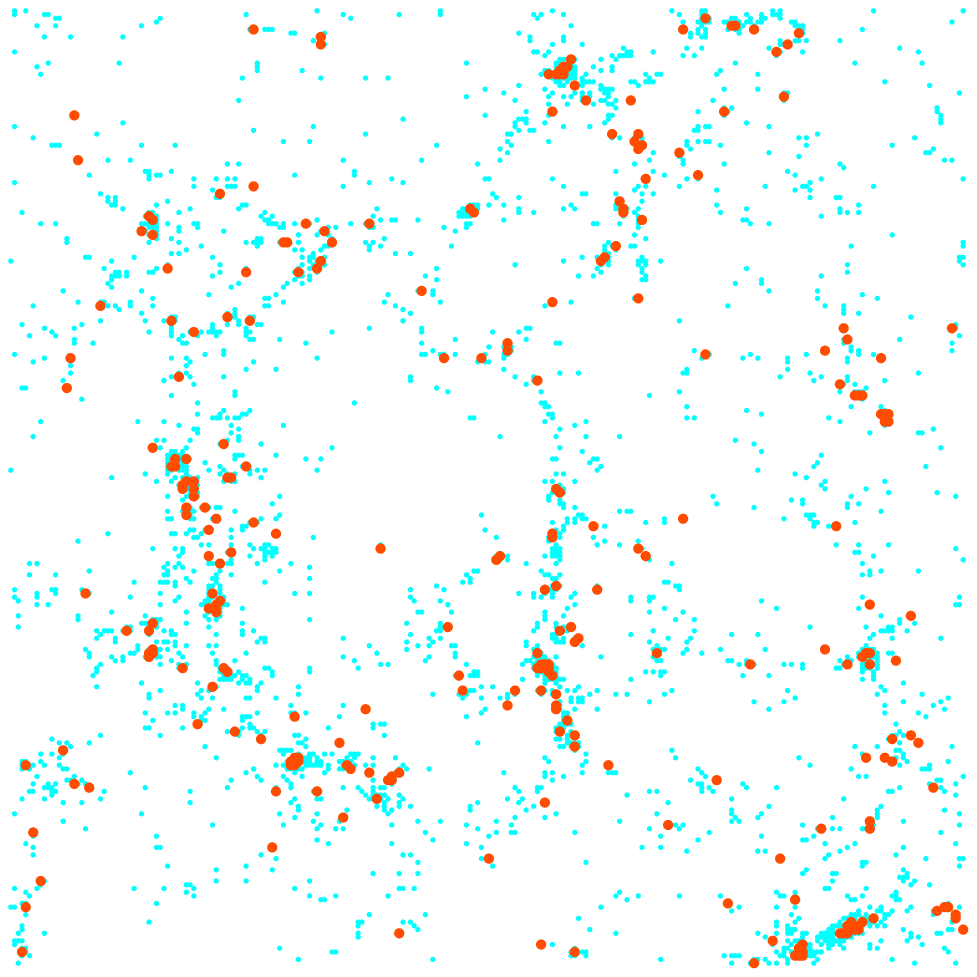}
\end{minipage}
\\
\begin{minipage}{7.5cm}
\includegraphics[width=7.5cm]{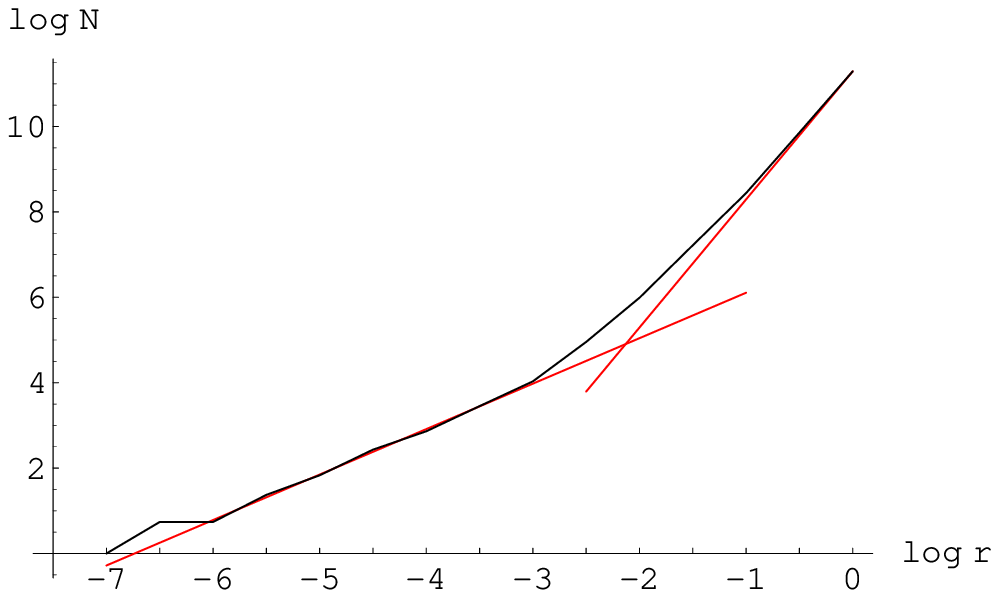}
\includegraphics[width=7.5cm]{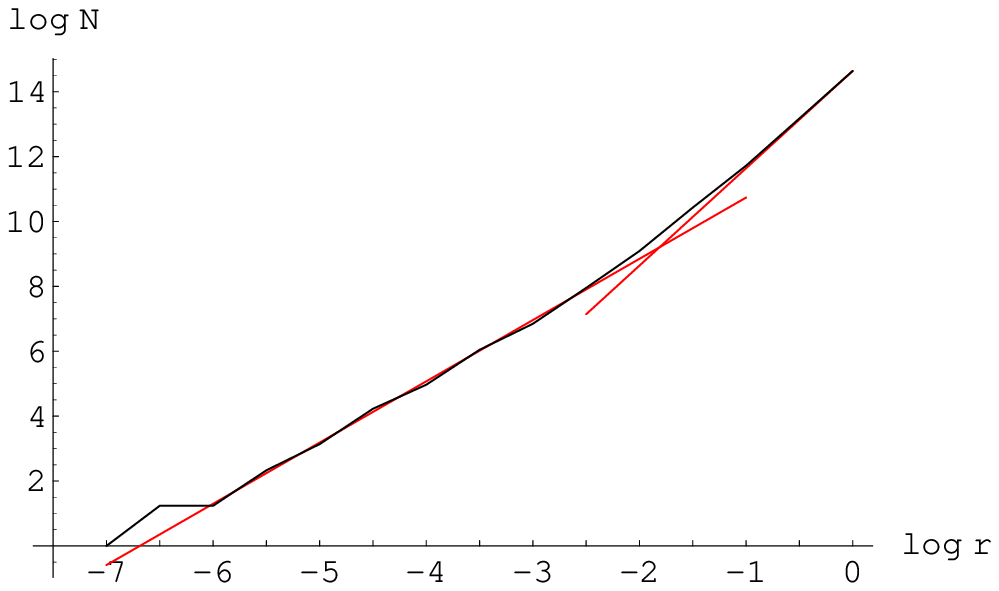}
\end{minipage}
\caption{Selection of two halo populations from the GIF2 $N$-body
simulation: heavy haloes with 750 to 1000 particles and light haloes
with 100 to 150 particles.  (Top) $1/8$-thick slice showing heavy
haloes in red and light haloes in blue.  (Down) Number-radius relation
for each halo population, showing fractal dimensions $1.1$ and $1.9$,
respectively, and a transition to homogeneity in both (logarithms are
to base $2$ and the total size is normalized to unity).}
\label{figs2}
\end{figure}

Results of the analysis appear in Fig.~\ref{figs2}: on the top, there
is a $1/8$-thick slice that shows the aspect of both halo populations;
below we have log-log plots of the number-radius functions for the
respective spatial distributions. They have similar scaling ranges,
but corresponding to quite different fractal dimensions, namely, $1.1$
and $1.9$.  The scale of transition to homogeneity, according to Eq.\
(\ref{N-r0}), is $r_0 \simeq 1/4$.  Intermediate mass values ($\sim
300$) yield intermediate dimensions but similar $r_0$.

\subsection{The structure of voids}
\label{voids-GIF2}

In our definition of void, in Sect.\ \ref{voids}, we outlined how
coarse-graining affects the structure of voids.  As said above, in
$N$-body simulations, halos only arise in the nonlinear stage, as they
concentrate particles from other regions that become {\em
voids}. Moreover, these voids are {\em totally empty}, because a cell
with less than one particle has to be empty.  However, an improvement
in mass resolution can be achieved with a relatively simple numerical
technique: re-simulation of selected regions. Indeed, re-simulation of
voids with higher resolution unveils new structure, similar to the
structure with the initial resolution (Gottl\"ober et al, 2003). This
new structure includes small halos. In the words of Gottl\"ober et al
(2003), ``the haloes are arranged in a pattern, which looks like a
miniature Universe''.  This type of self-similar structure corresponds
to a multifractal.

\begin{figure}
\includegraphics[width=7.5cm]{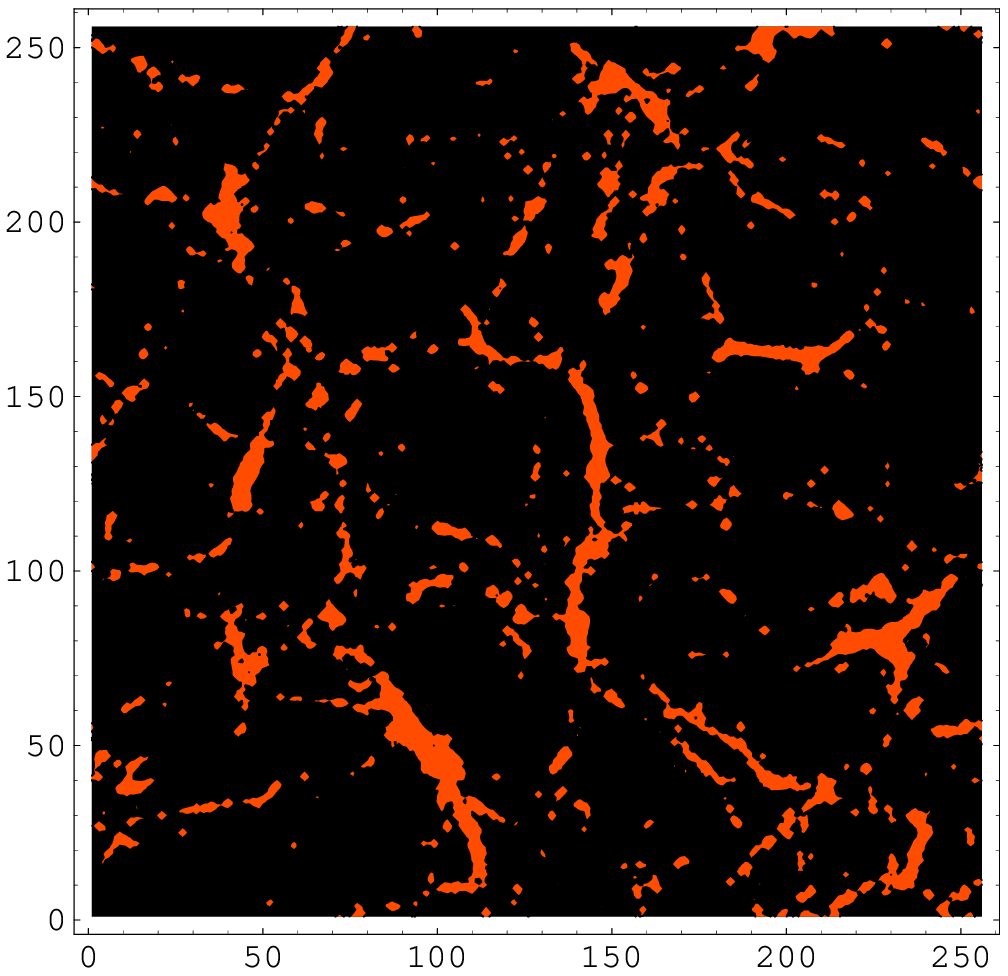}
\caption{A section of the isodensity contour of the smoothed density
field at the density corresponding to $\a = 3$. The dark area
corresponds to mass depletions with $\a = 3$ and the light area
corresponds to mass concentrations with $\a = 3$, namely, to halos.
The boundary is a very complex curve, with dimension larger than one.}
\label{figs3}
\end{figure}

In our analysis of the $\L$CDM GIF2 simulation, we have taken $\d_H =
256$, so the smallest halos have about 4 particles and
cells with fewer particles belong to voids. We note that the boundary
of voids cannot have a good pictorial definition in this situation.
To appreciate the complicated geometry of this boundary, we have
plotted in Fig.~\ref{figs3} a smoothed version of a two-dimensional
section of it (obtained by means of isodensity contours).

The voids displayed in Fig.~\ref{figs3} look quite empty.  However,
some particles lie in them. Moreover, re-simulation of a region in
them with higher resolution (as in Gottl\"ober et al, 2003) would
surely unveil new structure similar to the structure already formed by
halo clusters in Fig.~\ref{figs3}, but in smaller size. That is to
say, many small halo clusters would pop out in the voids. At the same
time, if the re-simulation were extended to the roundish small halo
clusters already displayed in Fig.~\ref{figs3}, they would become less
round and more crumpled, like the large ones.

\subsection{Multifractal spectrum}
\label{MF-spec}

Finding fractal distributions of halos with definite scaling exponents
constitutes a proof of multifractality.  However, we must also apply
the standard method of multifractal analysis, namely, the calculation
of $q$-moments and, hence, the multifractal spectrum, according to
Eqs.\ (\ref{tauq}) and (\ref{MFspec}).  Since $q$-moments have no well
defined scaling, $\tau(q)$ is not well defined. Nevertheless, it can
be computed at any given scale, understanding Eq.\ (\ref{tauq}) as a
an equation at the given scale rather than a limit.  Then, we can
study how the multifractal spectrum given by Eqs.\ (\ref{Leg}) and
(\ref{MFspec}) changes across the scales.

\begin{figure}
\includegraphics[width=7.5cm]{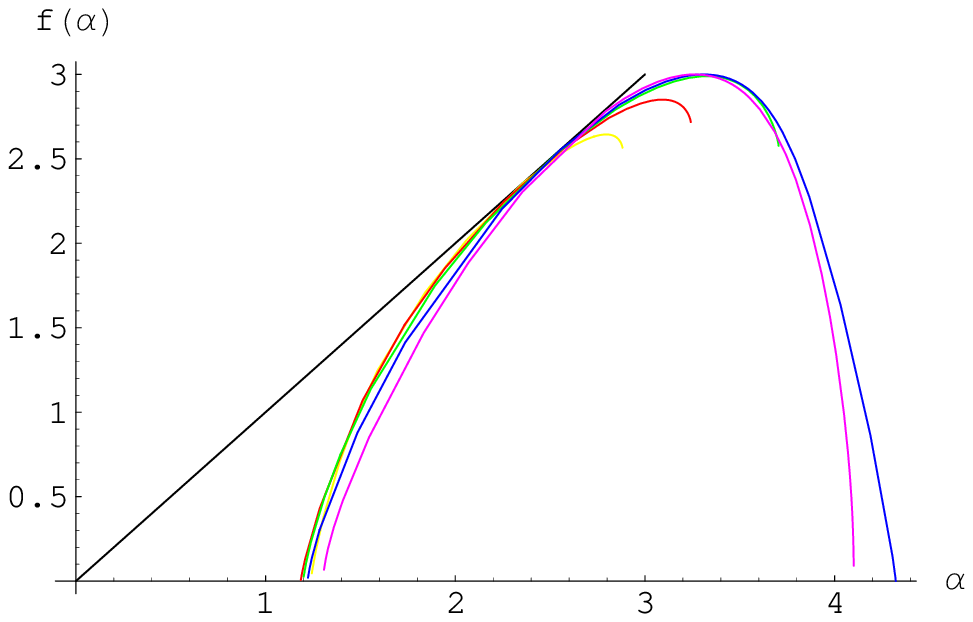}
\caption{Full multifractal spectrum $f(\a)$ for 
$\log_2\d = -9,-8,-7,-6,-5$
(yellow, red, green, blue, magenta). We have also plotted the diagonal
line to show how the multifractal spectrum is placed under it and
touches it at the measure's concentrate dimension $\a \simeq 2.5$.  }
\label{fig-MFspec}
\end{figure}

Therefore, we have computed the multifractal spectrum from $q$-moments
for various $\d_n = 2^{-n}$, namely, $\log_2 \d = -9,\ldots,-5$, and
plot the results in Fig.\ \ref{fig-MFspec}.  We can see that the
multifractal spectrum is stable in the range of $\a$ where all the
curves overlap. For the decreasing part of $f(\a)$, which corresponds
to negative $q$, we have only two curves. Their agreement is
sufficient for smaller $\a$ but they separate for larger $\a$ ($\a >
4$). This was to be expected, for these values correspond to the most
depleted voids, which are not sampled.

The collapse of the $f(\a)$ curves along such a considerable range of
$\a$ and for a considerable range of scales is surprising indeed
(compare with the log-log plot of $M_2(\d)$ in Fig.\ \ref{fig_M2}).
Furthermore, the stability of $f(\a)$ contrasts with the change of the
halo mass function from power law to lognormal that takes place over
the smaller scales in Fig.\ \ref{fig-MFspec}.  In this regard, our
conclusion is that the crucial condition in the calculation of the
multifractal spectrum by the method of moments has some sort of
self-consistency; namely, the condition that only singularities with
exponent $\a(q)$ contribute to the $q$-moment (Sect.\
\ref{moment-scaling}), allows one to {\em extrapolate} the
multifractal spectrum to small scales.

Here is a number of consequences that can be directly extracted 
from the coarse multifractal spectrum, such as is displayed 
in Fig.\ \ref{fig-MFspec}:
\begin{itemize} \itemsep 0mm
\item The entropy dimension is $\simeq 2.5$. This means that this
multifractal is concentrated in a set which is not very far from being
homogeneous (dimension three).
\item The largest fractal dimension is actually three, so  
this is the dimension of the measure's support.
In conjunction with
statistical homogeneity, this essentially
means that there are no totally empty voids in the full dark matter
distribution (see the comments in Sect.\ \ref{voids-GIF2}).
\item $\a(q=0) \simeq 3.3 > 3.$ This value corresponds again to the
multifractal support. We remark that it is formed by mass depletions,
that is, such that they belong to voids. Therefore, voids (though
{\em not} totally empty) dominate the distribution (see also
Fig.~\ref{figs3}).
\item The set of regular points with $\a = 3$ (non-vanishing density),
namely, the boundary of voids, turns out to have $f(\a) \simeq 2.9$.
This fractal dimension is very close to three.  This does not
necessarily mean that each piece of surface is very crumpled, because
the fractal dimension is also a measure of fragmentation (Mandelbrot,
1977). Therefore, its large dimension surely results from the
proliferation of small halo clusters that fill the voids. This type of
geometry can be inferred from Fig.~\ref{figs3}.
\end{itemize}

\section{Conclusions}

We have shown that a statistically self-similar multifractal model
explains the relevant features of the geometry of the large scale
structure of matter.  We have described this multifractal model as
consisting of singular mass concentrations and regular mass depletions
(plus a boundary between both).  Hence, we have proposed to identify
mass concentrations with halos and to identify mass depletions as
belonging to voids.  The identification of halos with mass
concentrations implies, in the scale-invariant limit, that they have
an {\em average} power-law profile, although they can actually be very
anisotropic.  In addition, halos are clustered with a power-law form
of its two-point correlation functions.  In particular, every
population of mass concentrations with equal strength (singularity
exponent) has a definite fractal dimension.

We have emphasized that scale invariance, in fact, forbids one to
distinguish the mass distributions of matter inside halos from the
distribution outside. That is to say, scale invariance actually
deprives halos of meaning as entities with a given size (or mass), so
they can only be classified by their {\em strength}
exponent. Therefore, halos can only be given a size in connection with
the breakdown of scale invariance.  In this connection, the number of
halos becomes finite and their average profile changes.  In
cosmological $N$-body simulations, we have argued that it is most
convenient to use the {\em uniform} coarse-graining length given by
the discretization scale (the linear size of the volume per particle).

The uniform size of halos in our multifractal model can be interpreted
in terms of standard coarse multifractal analysis.  In $N$-body
simulations, which use equal-mass particles, coarse multifractal
analysis becomes equivalent to the method of counts-in-cells.  Each
halo is formed by a variable number of particles, a number which can
be small. Each population of halos with (about) the same number of
particles forms a monofractal with a definite dimension.  We have
analysed, in the Virgo $\L$-CDM GIF2 simulation, the mass function of
halos and their distribution.  The distributions of the different
populations of halos show well-defined scaling ranges, corresponding
to different scaling dimensions, and shows a common scale of
transition to homogeneity.  We have selected two representative
populations of heavy or light halos, resulting in fractal dimensions
$D = 1.1$ or $D = 1.9$, respectively.  Note that these two values span
the usual range of fractal dimensions measured in the galaxy
distribution. The conclusion of our analysis of the GIF2 simulation is
that scale invariance holds in the nonlinear regime reproduced in
$N$-body simulations and that it can be measured, contrary to the
conclusion of Jenkins et al (1998).

The scale of transition to homogeneity in the fractal distributions of
halos of the GIF2 simulation is, in physical units, $r_0 \simeq 14\,
h^{-1}$ Mpc. Although this value corresponds to a dark-matter
simulation, it roughly agrees with standard values found in the
galaxy distribution.

The fractal distributions of halos with different dimensions are
biased with respect to the full particle distribution, which has fixed
mass-radius dimension $D(2)$.  Of course, this bias is different from
the usual linear bias assumed in cosmology, which only affects the
{\em amplitude} of the two-point correlation function: it is a bias
that affects the exponent, that is to say, the fractal dimension. We
may properly call this kind of bias {\em nonlinear bias of scaling
type}.

Mass depletions, belonging to voids, are very undersampled in $N$-body
simulations: by definition, depleted halo-size cells have no
particles; or they have a few of them, at the most, if we define the
halo size to be somewhat larger than the discretization scale. So the
few particles in voids are a gross representation of the small halos
that one would find there at a higher resolution. In this regard, our
conclusion is similar to that of Kuhlman, Melott \& Shandarin (1996).
However, self-similarity allows us to draw further conclusions
regarding the distribution of matter in voids, since the properties of
the distribution can be obtained by coarse multifractal analysis with
larger cell sizes.

As to the r\^ole of correlation moments, multifractals are defined by
the full set of moments $M_q(r)$ for $-\infty < q < \infty$ (or the
PDF moments $\mu_q(r)$ plus $M_0(r)$), whereas monofractals are
defined only by $M_2(r)$.  In particular, monofractals fulfill the
hierarchical relation $M_q \sim M_2^{q-1}$, for any $q$.  Thus, the
singular {\em and} heterogeneous nature of multifractals implies that
the moments with $q = 2,3, \ldots$ are not sufficient.  In fact, most
important are $q = 0,1$, associated to the the measure's support and
concentrate, respectively.  Larger $q$'s are also relevant, in
particular, $q=2$, but the information gathered from large-$q$ moments
only applies to the strongest concentrations (most massive halos).
The negative-$q$ moments are related to the distribution of matter in
voids. Therefore, they are affected by the undersampling of
voids. However, if the scaling range is sufficiently large, these
negative-$q$ moments can be calculated (approximately), as well as the
part of the multifractal spectrum given by them.

Regarding $q$-moments of the full mass distribution in $N$-body
simulations, we have seen that they scale poorly. In contrast, the
analysis in terms of fractal populations of equal-mass halos
by their number-radius relation (or two-point correlation function
of their positions) provides good scaling, suitable for calculating
the fractal dimension of each population.  The reason for the poor
scaling of $q$-moments is that the one-to-one correspondence between
mass concentrations (or depletions) and $q$-moments given by the
function $\a(q)$ is only approximate when the scaling range is
limited. If several concentration strengths contribute to the same
$q$-moment, their mixing spoils the scaling of that $q$-moment.  We
have discussed in Sect.\ \ref{counts} how this mixing of halo
populations takes place.

In particular, the various halo populations contribute evenly to $M_2$
and, hence, to the full two-point correlation, altering its
scaling. Therefore, it is natural that Jenkins et al (1998) observed
deviations from scaling in their analysis of $N$-body simulations. It
is also natural that Zehavi et al (2004) observe deviations from
scaling in their analysis of the Sloan Digital Sky Survey: if one
associates galaxies to halos, their distribution is likely
multifractal, so the mixing of populations takes place as well.  In
fact, Zehavi et al (2005) and Tikhonov (2006) have found that the
slope $\g$ of the log-log plot of the two-point correlation decreases
with luminosity, which agrees with a multifractal distribution.

In the GIF2 simulation, at the scale of the halo size $\d_h$, the mass
function is well described by a power law, so that it corresponds to a
linear multifractal spectrum and the distribution is actually {\em
bifractal}.  To be precise, we find $N(m) \sim m^{-2}$, very
accurately. Therefore, the corresponding bifractal fulfills $\a =
f(\a)$ and is an even bifractal, in which the measure is uniformly
concentrated (not loaded toward either the more or the less singular
end, as explained in Sect.\ \ref{linearMF}).

Thus, the bifractal mass function furnished by equal-size halos must
be distinguished from the Press-Schechter mass function, which (apart
for the large-mass cutoff) is equivalent to a linear mutifractal
spectrum concentrated on the singular end, that is to say, on the
heavy halos. Note that Eq.\ (\ref{P-S_MF}) shows that a mass function
$N(m) \sim m^{-2}$ is associated with the spectral index $n = -3$,
which is just beyond the allowed range. Of course, equal-size and
Press-Schechter halo mass functions were expected to differ.  However,
it is remarkable that both have a power-law range and that the
constant exponent that appears with our definition is a limit case of
the Press-Schechter mass function. In any event, we must point out
that the exponent $-2$ for equal-size halos is {\em independent of the
initial power spectrum}, in accord with the nonlinear nature of the
multifractal.

The mass function of halos is bifractal (a power law), but the
calculation of the multifractal spectrum through moments for cell
sizes larger than $\d_h$ shows convergence towards a convex,
lognormal-like spectrum (Sect.\ \ref{MF-spec}).  We conclude that, at
the halo-size scale $\d_h$, the mass discretization does not allow for
a good sampling of halos and allows for almost no sampling of voids.
Poorly sampled halos mainly reflect the measure's
concentrate. Therefore, the power law is merely a linear approximation
to the full mass function near the measure's concentrate.  In this
regard, let us remark that, in a multifractal that is scale invariant
on ever decreasing scales, the {\em total} mass concentrates in the
set such that $\a = f(\a)$. Paradoxically, mass concentrations with
larger $\a$ are too few to hold any mass, even all together. On the
other hand, mass concentrations with smaller $\a$ (or mass depletions)
are abundant, but their total mass vanishes nonetheless.

\acknowledgments
I am grateful to Liang Gao for kindly supplying me with the GIF2 data.
My work is supported by the ``Ram\'on y Cajal'' program and by grant
BFM2002-01014 of the Ministerio de Edu\-caci\'on y Ciencia.


\begin{thebibliography}{0}

\bibitem{Alimi}
Audit E., Teyssier R. \& Alimi J.-M., 1997,
A\&A, 325, 439

\bibitem{Aurell}
Aurell E., Frisch U., Noullez A. and Blank M., 1997, 
Journ.\ Stat.\ Phys.\ 88, 1151

\bibitem{Bal-Schaf}
Balian R. \& Schaeffer R., 1988, ApJ, 335
, L43

\bibitem{Bal-Schaf1}
Balian R. \& Schaeffer R., 1989-A, A\&A, 226, 1--29

\bibitem{Bal-Schaf2}
Balian R. \& Schaeffer R., 1989-B, A\&A, 226, 373--414

\bibitem{BMP}
Botaccio M., Montuori M. \& Pietronero L., 2004, 
Europhysics Letters, 66, 610

\bibitem{Borga}
Borgani S., 1993, MNRAS, 260, 537

\bibitem{Colom}
Colombi, S., Bouchet, F.R. and Schaeffer, R., 1992,
A\&A, 263, 1

\bibitem{CoorShet} Cooray A. \& Sheth R., 2002, Phys. Rep. 372, 1

\bibitem{FoF} Davis M., Efstathiou G., Frenk C. and White S.D.M., 
1985, ApJ, 292, 371.

\bibitem{multif1}
Dom{\'\i}nguez-Tenreiro R. \& Mart{\'\i}nez V.J., 1989, ApJ 339, L9

\bibitem{Falc} Falconer K., 2003, Fractal geometry --- Mathematical 
Foundations and Applications, Second Edition, John Wiley \&
Sons, Chichester, UK

\bibitem{Frisch}
Frisch U., 1995, Turbulence: The Legacy of A.N. Kolmogorov,
Cambridge University Press.

\bibitem{I} Gaite, J., 2005-A,
Europhysics Letters 71, 332--338 

\bibitem{I2} Gaite, J., 2005-B,
Eur.\ Phys.\ Jour.\ B, 47, 93 

\bibitem{I3} Gaite, J., 2006, Physica D 223, 248

\bibitem{Gaite} 
Gaite, J. and Manrubia S. C., 2002,
MNRAS, 335, 977

\bibitem{GIF2} 
Gao L., White S.D.M., Jenkins A., Stoehr F. \& Springel V., 
2004, MNRAS, 355, 819

\bibitem{Gott} Gottl\"ober, S., Lokas, E.L., Klypin, A. \& Hoffman, Y.,
2003, MNRAS, 344, 715

\bibitem{Halsey}
Halsey T.C., Jensen M.H., Kadanoff L.P., Procaccia I. \& Shraiman B.I.,
1986, Physical Review A, 33, 1141

\bibitem{Harte} Harte D., 2001, Multifractals: theory and applications, 
Chapman \& Hall

\bibitem{Virgo}
Jenkins, A., Frenk, C. S., Pearce, F. R., Thomas, P. A., Colberg,
J. M., White, S. D. M., Couchman, H. M. P., Peacock, J. A.,
Efstathiou, G., and Nelson, A. H., 1998,
ApJ 499, 20

\bibitem{Jones2} Jones B.J., Coles P. and Mart\'{\i}nez V., 1992, 
MNRAS, 259, 146

\bibitem{Jones} Jones B.J., Mart\'{\i}nez V., Saar E. and Einasto J., 
1988, ApJ, 332, L1

\bibitem{Jones-RMP} 
Jones B. J., Mart\'{\i}nez V. J., Saar E. and Trimble V., 2004,
Rev.\ Mod.\ Phys.\ 76, 1211

\bibitem{KMS} Kuhlman B., Melott A.L. and Shandarin S.F., 1996, ApJ, 470, L41

\bibitem{Lee-Shan} Lee J. and Shandarin S.F., 1998, ApJ, 500, 14--27

\bibitem{Mandel} Mandelbrot B.B., 1977,
The fractal geometry of nature, 
W.H. Freeman and Company, NY

\bibitem{multif2}
Mart{\'\i}nez, V.J., Jones, B.J., Dom{\'\i}nguez-Tenreiro, R.
\& van de Weygaert, R., 1990, ApJ, 357, 50

\bibitem{McC-Silk}
McClelland J. \& Silk J., 1977, ApJ, 217, 331

\bibitem{Mo-W}
Mo, H.J. and White, S.D.M., 1996, 
MNRAS, 282, 347

\bibitem{Mur}
Murante G., Provenzale A., Spiegel E.A. \& Thieberger R., 1997,
MNRAS 291, 585

\bibitem{Pari-F} Parisi G. \& Frisch U., 1985, On the singularity
structure of fully developed turbulence, in {\em Turbulence and
Predictability in Geophysical Fluid Dynamics, Proc. Intnal. School of
Physics `E.\ Fermi', 1983, Varenna (Italy),} 84--87, eds.\ M.\ Ghil,
R.\ Benzi and G.\ Parisi, North-Holland, Amsterdam

\bibitem{Pee}
Peebles P.J.E., 1974, A\&A, 32, 197

\bibitem{Pee2} Peebles, P.J.E., 1980,
The large-scale structure of the universe,
Princeton U.\ Press, Princeton, NJ

\bibitem{Piet}
Pietronero L., 1987, Physica A 144, 257--284

\bibitem{Shan} Shandarin S.F., Sheth J.V. and Sahni V., 2004,
MNRAS, 353, 162

\bibitem{She} She Z., Aurell E. and Frisch U., 1992, 
Commun.\ Math.\ Phys., 148, 623

\bibitem{Sheth-J}
Sheth, R.K. and Jain, B., 1997, 
MNRAS, 285, 231

\bibitem{Sheth-Tor}
Sheth, R.K. and Tormen, G., 1999, 
MNRAS, 308, 119

\bibitem{Sornette} Sornette, D., 2000,
Critical phenomena in natural sciences, Springer-Verlag

\bibitem{Piet-Marti}
Sylos Labini F., Montuori M. and Pietronero L., 1998,
Phys.\ Rept., 293, 61

\bibitem{Tikho}
Tikhonov A., Correlation Properties of Galaxies from the Main Galaxy 
Sample of the SDSS Survey,
preprint {\tt astro-ph/0610643}

\bibitem{Vala} Valageas P., 1999, A\&A, 347, 757

\bibitem{Vala2} Valageas P., Lacey C. and Schaeffer R., 2000,
MNRAS, 311, 234--250

\bibitem{Valda}
Valdarnini R., Borgani S. \& Provenzale A., 1992, 
ApJ, 394, 422

\bibitem{V-Frisch} Vergassola M., Dubrulle B., Frisch U. \& Noullez A.,
1994, A\&A 289

\bibitem{Yepes}
Yepes G., Dom{\'\i}nguez-Tenreiro R. \& Couchman, H.P.M., 1992,
ApJ, 401, 40

\bibitem{Zehavi_1}
Zehavi I., et al, 2004, ApJ, 608, 16

\bibitem{Zehavi_2}
Zehavi I., et al, 2005, ApJ, 630, 1

\bibitem{Zinne} Zinnecker H., 1984, MNRAS, 210, 43

\bibitem{Zipf}
Zipf G.K., 1949, 
{Human behavior and the principle of least effort},
Addison-Wesley Press Inc., Massachusetts


\end{thebibliography}
\end{document}